\documentclass[journal]{IEEEtran}

\ifCLASSINFOpdf

\fi
\usepackage{authblk}
\usepackage{amsmath}
\usepackage{epsfig}
\usepackage{array}
\usepackage{subfig}
\usepackage{graphicx,amssymb}
\usepackage{capt-of}
\usepackage{algorithm}
\usepackage{algorithmic}
\usepackage{setspace}
\usepackage{epstopdf}
\usepackage[export]{adjustbox}
\usepackage{subfig}
\usepackage{graphicx}
\usepackage{stfloats}
\usepackage{filecontents,lipsum}
\usepackage[noadjust]{cite}
\usepackage{amsthm}

\theoremstyle{definition}

\newtheorem{proposition}{Proposition}

\def\bt{{\mathbf t}}
\def\bz{{\mathbf z}}
\def\bd{{\mathbf d}}
\def\balpha{{\boldsymbol{\alpha}}}
\def\bgam{{\boldsymbol{\gamma}}}
\def\bB{{\mathbf B}}

\def\bx{{\mathbf x}}

\def\bz{{\mathbf z}}

\def\cT{{\mathcal T}}
\def\cS{{\mathcal S}}
\def\cN{{\mathcal N}}
\def\cC{{\mathcal C}}
\def\cP{{\mathcal P}}

\def\cE{{\mathcal E}}

\begin{document}
\title{Online Bayesian Inference of Diffusion Networks}
\author[1]{{\bf Shohreh Shaghaghian}}
\author[1]{{\bf Mark Coates}}  
\affil[1]{Department of Electrical and Computer Engineering, McGill University, Montreal}
\affil[ ]{\texttt{shohreh.shaghaghian@mail.mcgill.ca , mark.coates@mcgill.ca}}

\maketitle

\begin{abstract} 
  Understanding the process by which a contagion disseminates throughout a
  network is of great importance in many real world applications.
  The required sophistication of the inference approach depends on the type of
  information we want to extract as well as the number of observations
  that are available to us. We analyze scenarios in
  which not only the underlying network structure (parental
  relationships and link strengths) needs to be detected, but also the
  infection times must be estimated. We assume that our
  only observation of the diffusion process is a set of time series,
  one for each node of the network, which exhibit changepoints when an
  infection occurs. After formulating a model to describe the
  contagion, and selecting appropriate prior distributions, we seek to find the set of
  model parameters that best explains our observations. Modeling the problem
  in a Bayesian framework, we exploit Monte Carlo Markov Chain,
  Sequential Monte Carlo, and time series analysis techniques to
  develop batch and online inference algorithms. We evaluate the performance
  of our proposed algorithms via numerical simulations of synthetic network contagions and analysis
  of real-world datasets.
\end{abstract}

\IEEEpeerreviewmaketitle

\section{Introduction} \label{intro} 

The notion of transmission of some sort of behavioral change from one
agent to another exists in many phenomena around us. A very tangible
example is the spread of hashtags in social networks caused by
the influence that users have over one another
(e.g.~\cite{guille2013information,yang2010modeling,
  Wang2012Sparse,gomez2011uncovering, gomez2012inferring}). Other
examples include the propagation of distortions (which are caused by
external events) in stock returns of different assets in the stock
market (e.g.~\cite{ozsoylev2014investor,ahern2013network}), and the
outbreak of a contagious disease in different geographic regions
(e.g.~\cite{sefer2015convex}). What qualifies all of these inherently
different phenomena to be studied as a diffusion process is the
commonality of three main components: {\em Nodes}, i.e., the set of separate agents;
{\em Infection}, i.e., the change in the state of a node that can be
transferred from one node to the other; and {\em Causality}, i.e., the
underlying structure based on which the infection is transferred
between nodes. The term {\em cascade} is often used to refer to the
temporal traces left by a diffusion process.

Despite the common components mentioned above, diffusion scenarios can
be studied for different purposes~\cite{guille2013information}. In
some cases (e.g. \cite{ozsoylev2014investor,ahern2013network}), the
goal is to identify the most influential nodes of the network. The
main concern in other studies (e.g. \cite{shamma2011peaks}) is to
detect the popular topics that have diffused throughout a
network. In this paper, we pursue the goal of a third group of studies
(e.g. \cite{gomez2011uncovering, gomez2012inferring}), i.e., to infer
the causality component of a diffusion process using observations
related to the cascades. In other words, we aim to exploit the
available evidence to infer the path that an infection has traversed in
order to reach an arbitrary node. Inferring this path not only gives
us valuable insight about the existing dynamics between the nodes of the
network, it also helps us predict, expedite, or prevent the spread of
future infections.

\begin{figure*}[bt] \centering \includegraphics[width=\textwidth,
  height=0.42\textwidth]{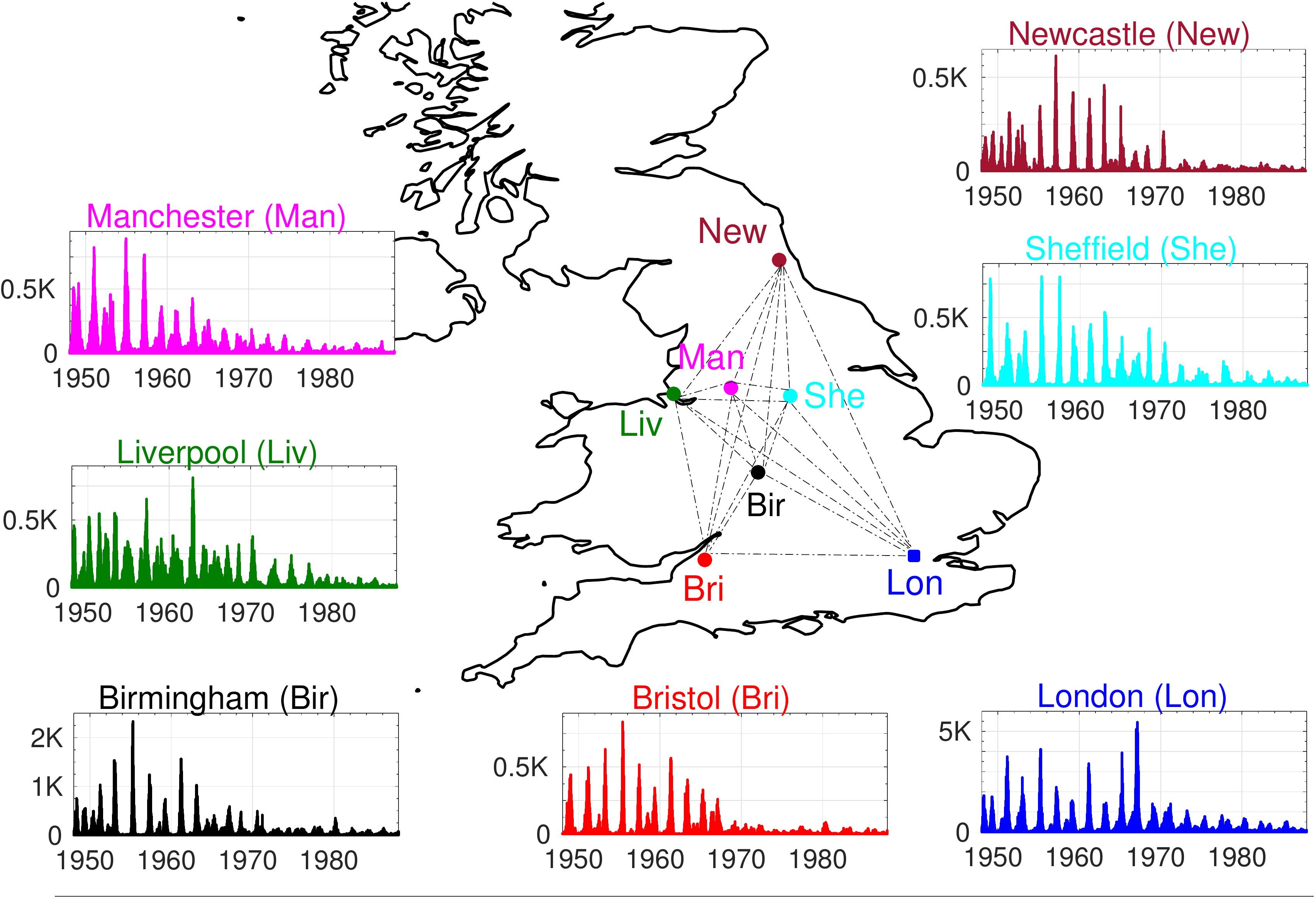} \caption{ Weekly Reports of
    Measles and Chickenpox in 7 Major Cities of England and Wales:
    London (Lon), Bristol (Bri), Liverpool (Liv), Manchester (Man),
    Newcastle (New), Birmingham (Bir), Sheffield (She)}
  \label{fig:map1} \end{figure*}

A major factor that differentiates network inference methodologies is
the type of available observations and the amount of useful
information that can be extracted from them. In most of the social
network inference scenarios, for example, the cascade trace is
directly observable. The moment of time at which an arbitrary node
shows infection symptoms can be easily specified and identified. The
majority of the available network diffusion inference literature
(e.g.~\cite{yang2010modeling,
  Wang2012Sparse, gomez2011uncovering,gomez2013structure,
  farajtabar2015coevolve, embar2014bayesian,gomez2012inferring}) focuses
on such frameworks where the cascades are directly and perfectly
observed. However, there are other scenarios in which we cannot easily
determine when nodes become infected. An example of such a scenario is
presented in Figure \ref{fig:map1}. Having access to the number of
weekly reported cases of measles and chickenpox in seven major cities
of England and Wales for almost 40 years\footnote{Available at
  https://ms.mcmaster.ca/bolker/measdata.html}, we can observe that
different regions become infected at different points in time. The
infection then dies away (the region returns to a ``susceptible'' state).
However, we cannot easily pinpoint the exact week when a region becomes
infected. Therefore, more sophisticated inference methods are required
for cases with limited or indirect observations.
\cite{sefer2015convex,farajtabar2015back,amin2014learning,
  lokhov2015efficient} have studied diffusion processes in which
the cascade trace is not directly observable or is partially missing.
In~\cite{amin2014learning, lokhov2015efficient,
  farajtabar2015back} it is assumed that a portion of the cascade
data is directly observable and the authors propose techniques to infer the
causality structure from this portion. In~\cite{sefer2015convex},
the cascade trace is unavailable, but other observable properties of
the cascade are used to infer the causality structure.
Although these approaches can identify how an infection has traversed
the network (in the example of Figure~\ref{fig:map1}, which region is
primarily responsible for infecting another), they sometimes do not
provide all of the information required to take appropriate action to
mitigate or expedite the spread of an infection. In the example of a
highly contagious disease, we may choose to control transportation and movement
between regions or implement stricter health checks. In the example of
the stock market, we may choose to invest in a stock that is likely to
be affected (infected) soon by an external disruptive influence. To
implement such strategies, it is important to estimate when nodes have
become infected and to infer model parameters that allow us to
construct predictions of when future infections will occur.

In this paper, we contribute to the available literature by developing
batch and online algorithms that simultaneously infer the causality
structure and estimate the unobserved infection times. We assume that
the only observation we have from each node is a time series whose
characteristics provide an indication of the behavioral changes caused by
receiving the infection. The paper is organized as follows. In the
next section, we briefly review related work. In Section \ref{model},
we describe our system model and formulate the diffusion problem. We
present our proposed batch and online inference approaches
respectively in Sections \ref{batch} and \ref{online}. We evaluate the
performance of our suggested methods using both synthetic and real
world datasets and present the simulation results in Section
\ref{res}. The concluding remarks are made in Section \ref{Conclu}.
  \begin{table*} \centering
  \setlength{\extrarowheight}{2pt}
  \begin{tabular}{lll}
    	\hline
    {\bf Symbol(s)} &{\bf Expression(s)} &{\bf Definition(s)} \\
	\hline
	$\cN$ & $\{1,\dots,N\}$ &Set of all the $N$ nodes in the network\\
	$\pi^i$ & - &Set of potential parents for node $i$\\
	$\kappa_{ij},\theta_{ij}$ & - &Gamma distribution hyperparameters for link $ij$ \\
	$N_T,M,N_B$ & $N_T=M\times N_B$& Length of observed data, Number of blocks, Length of each block\\
	$\bgam_1,\bgam_2$ & $\bgam_j=(\gamma_j^1,\dots,\gamma_j^N)$, $j=1,2$ &Hyperparameters of observed data before and after being infected\\	
	$\bB_b$ & $[M(b-1)+1,\dots,Mb]^T$ &Vector of time indices in block $b$  \\
	$\bd_{\bB_b}$ & $\{ \bd_{\bB_b}^1,\dots, \bd_{\bB_b}^N \}$ &Set of data received in block $b$ \\ 
	$\bz_b$ &$[z_b^1,\dots,z_b^N]^T$ & Random vector of parents in block $b$ \\
	$\bt_b$&$[t_b^1,\dots,t_b^N]^T$ & Random vector of infection times in block $b$\\
	$\balpha_b$&$[\alpha_b^1,\dots,\alpha_b^N]^T$  & Random matrix of link strengths in block $b$\\
	$\bx_b$&$(\bz_b,\bt_b,\balpha_b)$ & Infection parameters  in block $b$\\
	$p_b^i$&$1-e^{-\alpha_b^{iz_b^i}}$& Parameter of geometric distribution in block $b$\\	
	$\bx_b^{MAP}$ &$(
        \hat{\bz}_b,\hat{\bt}_b,\hat{\balpha}_b)$ & MAP estimates of infection parameters in block $b$\\
	$\cC_b^i$ & $\{k|z_b^k=i\}$&Set of node $i$'s children in block $b$\\
	$\pi_b^i$ & $\{k| k\in \pi^i, t_b^k\neq \phi, t_b^k < t_b^i\}$ &Set of node $i$'s potential parents in block $b$\\
	$\bz_b^{\overline{i}}$ & $[z_b^1,\dots,z_b^{i-1},z_b^{i+1},\dots,z_b^N]^T$ &Vector of parents except node $i$ in block $b$\\
	$\bt_b^{\overline{i}}$ & $[t_b^1,\dots,t_b^{i-1},t_b^{i+1},\dots,t_b^N]^T$ &Vector of infection times except node $i$ in block $b$\\
	$\balpha_b^{\overline{ij}}$ & $[\alpha_b^{i1},\dots,\alpha_b^{i(j-1)},\alpha_b^{i(j+1)},\dots,\alpha_b^{iN}]^T$ & Vector of node $i$'s link strengths except for link $ij$ in block $b$\\
	$\bx_b^m$ & $( \bz_b^m,\bt_b^m,\balpha_b^m )$ &The
        $m^{ \text{th}}$ sample from the distribution $f(\bx_b|\bd_{\bB_{1:b}})$ in block $b$\\
	$r_b$ & - &  Geometric parameter in the proposal for infection times in block $b$\\
	$\bt_b^{ML}$ & $[{t_b^{ML}}^1,\dots,{t_b^{ML}}^N]^T$ & Vector of maximum likelihood changepoints of $\bd_{\bB_b}$\\
	$N_{thin}, N_{burn}$ & - &Thinning and burn-in factors\\
	$N_{MCMC},N_{s}$& $N_{MCMC}=N_{s}\times N_{thin}+N_{burn}$ & Number of generated and stored particles\\
	\hline
	\end{tabular}
	\caption{Summary of Notations Used in the Paper}
	\label{ta:not}
\end{table*}
\section{ Related Work} \label{Related_Work}

In this section, we review the existing research most closely related to this
paper. We first discuss the studies that assume the cascades
(infection times) are perfectly observed and propose methods to infer
just the network structure (\ref{related1}). Then in \ref{related2},
we describe the few existing studies that have the same assumption as
this paper and clarify how this paper contributes to this body of
literature. The key assumption is that neither the network structure
nor the infection times are perfectly or directly observed. In
\ref{related3}, we survey methods for detecting the
moment of time at which the statistical characteristics of multiple time
series change. These methods do not involve any notion of an underlying diffusion network
that induces relationships between the time series. 

\subsection{Network Inference with Perfect Cascade Observation}\label{related1}

Most of the earlier work exploring diffusion network inference
techniques assumes that cascades are perfectly observed, i.e., the
infection times are exactly known. \cite{gomez2011uncovering} models
diffusion processes as discrete networks of continuous temporal
processes occurring at different rates. Given the infection times, the
goal is to infer the parental relationships and estimate the link strengths
that maximize the likelihood of the observed data. An algorithm called
NETRATE is developed to solve the convex optimization problem.
\cite{gomez2012inferring} uses a tree-shaped graph of the parental
relationships inferred from observed cascades of different contagions
to infer the complete set of edges of the graph (for example a
friendship graph in a social network). The proposed NETINF algorithm
is used to find the maximum likelihood graph conditioned on the
set of observed cascades.

\cite{gomez2013structure} uses the same setup as
\cite{gomez2011uncovering}, but it assumes that the underlying network
structure is not static and infection pathways change over time. A
stochastic convex optimization is employed to infer the dynamic
network and an online inference algorithm called INFOPATH is developed
to solve it. \cite{farajtabar2015coevolve} studies how the network
evolution affects the diffusion process and proposes a joint
continuous-time model to account for co-evolutionary dynamics between
these two processes.

\cite{embar2014bayesian} considers inferring the network structure as
an intermediate task and focuses on estimating joint properties of networks and
diffusion processes such as the node influence score of a contagion. The
network structure (parental relationships and link strengths) is
assumed to be hidden and the infection times are observed. A Bayesian
framework is used to calculate the expectation of the hidden parameters under
the posterior distribution. Instead of inferring the network
structure, \cite{yang2010modeling} and \cite{Wang2012Sparse} focus on
the global influence a node has on the rate of diffusion through the
network. The authors develop a linear influence model in which the
growth in the number of newly infected nodes is expressed as a function of the infection
times of the previously infected nodes. \cite{yang2010modeling} shows
that the influence function of each node can be estimated using a simple
least squares procedure by modelling it in a non-parametric way.
\cite{Wang2012Sparse} uses the same linear influence model, but introduces
sparsity in the estimated influence matrix and applies regularization
penalties to take into account the nodes' centralities. 

\subsection{Network Inference without Perfect Cascade Observation}\label{related2}

We now review the existing inference techniques for scenarios where
cascades are not perfectly observed. Assuming that the infection times
are only partially observed and the diffusion trace is incomplete,
\cite{farajtabar2015back} develops a two-stage framework to pinpoint
the infection source. After learning a continuous-time diffusion
network model based on the historical diffusion trace in the first
stage, the source of an incomplete diffusion trace is identified by
maximizing its likelihood under the learned model. Importance sampling
approximation is used to find optimal solutions. Using a four state
infection model (Susceptible, Exposed, Infected, and Recovered),
\cite{sefer2015convex} assumes that partially observed probabilistic
information about the state of each node is provided, but the exact
state transition times (infection times) are not observed. The underlying network is
inferred by minimizing the expected loss over all realizations of the
unobservable trace. The loss function is the negative log-likelihood of node state probabilities at each observed time point in a realization of infection times. Although the network structure can be detected using a convex optimization problem, the transition (infection) times are not
estimated.

\cite{amin2014learning} studies the theoretical learnability of tree-like graphs given the initial and final set of infected nodes.
The traces are defined as sets of unordered nodes and 
the authors strive to reconstruct the underlying network. Their proposed algorithm works by observing the relation that a particular vertex's infection has on the likelihood of infection at other locations in the tree. The goal
in~\cite{lokhov2015efficient} is to reconstruct the so-called node couplings using Dynamic Message Passing (DMP) equations. The authors assume that the cascade observations are only partially available and define coupling of nodes $i$ and $j$ as the probability that the infected node $i$ transmits the contagion to its susceptible neighbour $j$.
 
In this paper, we improve upon these existing methods by proposing
an approach to simultaneously infer the structure and cascade trace
of a diffusion process when the infection times are completely unavailable. The batch
inference approach presented in this paper (see Section \ref{batch})
was introduced in an earlier conference
paper~\cite{shaghaghian2016bayesian}, but here we include more
extensive experimental results and develop an online version of the
inference algorithm.

\subsection{Changepoint Detection}\label{related3}

Another sizeable, related body of literature addresses detecting abrupt changes
in the statistical structure of multiple time series. The
moments in time that divide time series into distinct homogeneous
segments are referred to as
changepoints. We refer the reader to~\cite{eckley2011analysis} for a
detailed discussion of the topic.

Most changepoint detection, or time series
segmentation, methods strive to detect single
and multiple changepoints in univariate
\cite{fearnhead2005exact,eckley2011analysis,killick2012optimal} or
independent multivariate \cite{matteson2014nonparametric} time
series. More closely related to our work is the approach in
\cite{xuan2007modeling}, which involves an underlying Gaussian graphical
model that captures the correlation structure between multivariate
time series. There is no notion of a diffusion process; the model
captures contemporaneous correlation structure. In this
paper, we strive to detect the changepoints of multiple time series in the
context of a background diffusion process that dictates when the
changepoints occur. We combine the indications of change
in the statistical characteristics of the observed cascades with the
causality relationships between nodes to infer the underlying
structure as well as the infection times (i.e., the changepoints).

\section{System Model}\label{model}

We consider a set of $N$ nodes $\cN=\{1,\dots,N\}$ that are exposed to
a contagion $C$. We assume that $C$ originates in a subset of nodes
and is transmissible to other nodes of the network. We denote the
moment of time at which node $i$ receives the contagion by $t^i$. When
node $j \in \cN$ transfers the contagion to node $i \in \cN$ for the
first time, we say $i$ is infected by $j$. In this case, node $j$ is
referred to as the parent of node $i$ and is denoted by $z^i$. In this
paper, we focus on the {\it Susceptible-Infected (SI)} scenarios in
which an arbitrary node $i$ is infected by the first node that
transfers the infection to it and never recovers afterwards.  Each
node $i$ has a set of potential parents $\pi^i$, but it is infected by
only one of them i.e. $z^i \in \pi^i$. The factor that determines
which member of $\pi^i$ transfers the infection to node $i$ is the
strength of the relationship between node $i$ and each of its
potential parents $j \in \pi^i$. We denote this link strength for
nodes $i$ and $j$ by $\alpha^{ij}$.
 The definitions of parents and candidate parents simply imply that
$\forall j \in \pi^i : t^j < t^i$ and $\forall j \notin \pi^i :
\alpha^{ij}=0$. Now that we have characterized both the diffusion
components and the cascades, we assign a directed, weighted graph to
the diffusion process. A directed edge $j\rightarrow i$ with weight
$\alpha^{ij}$ exists in this graph if and only if $z^i=j$ . We denote
this graph by $G=(\cN,\cE,\balpha_{N \times N})$ where $\cE$ is the
set of directed edges, and $\balpha=[\alpha^{ij}]_{N\times N}$ is the
weight matrix. Throughout the paper, we use $[\cdot]$ to denote
vectors and matrices, $\{\cdot\}$ to denote sets, and $(\cdot)$ to
denote finite ordered lists.

As mentioned in Section \ref{intro}, we focus on the scenarios where
neither the network structure (i.e., parental relations and link
strengths) nor the cascades (i.e., infection times) are directly
observed. We assume that the only observation we get from an arbitrary
node $i \in \cN$ is a discrete time signal of length $N_T$ denoted by
$\bd^i=(d^1_n,\dots,d_n^{N_T})$. We denote the set of all observed time
signals by $\bd=\{ \bd^1,\dots,\bd^N \}$.

In order to develop inference algorithms for the case where data
arrives in a streaming fashion or in batches, we consider a setting
where each node's data arrives in $N_B$ blocks of length $M$. Denote
the vector of time indices in block $b$ by $\bB_b$ i.e.
$\bB_b=[M\times(b-1)+1,\dots,M\times b]^T$ and the data in block $b$ for
all the nodes in the network by $\bd_{\bB_b}=\{
\bd_{\bB_b}^1,\dots,\bd_{\bB_b}^{N} \}$. 

We denote the parent for node $i$ at the end of the $b^{\text {th}}$
block by $z_b^i$. If no infection has occurred by the end of block
$b-1$, but it occurs during the $b^{\text{th}}$ block, then $z_{b-1}^i$ has a
null-value, i.e., $z_{b-1}^i=\phi$, whereas $z_{b}^i = j$ for some $j \in
\pi_i$. Likewise, the infection time for node $i$ and the link
strength associated with link $ij$ in block $b$ are respectively
denoted by $t_b^i$ and $\alpha_b^{ij}$. If no infection has occurred
then $t_b^i$ also has a null-value $\phi$.

The parameters at the end of the $b^{\text{th}}$ block are
denoted by $\bx_b=( \bz_b,\bt_b,\balpha_b)$ where
$\bz_b=[z_b^1,\dots,z_b^N]^T$, $\bt_b=[t_b^1,\dots,t_b^N]^T$, and
$\balpha_b=[\alpha_b^{ij}]_{N \times
  N}$. While the set of potential parents $\pi^i$ is the same for all
the blocks, two sets $\cC_b^i$ and $\pi_b^i$ are defined as follows
for each block $b$. 
\begin{equation*}
  \cC^i_b=\{k|z_b^k=i\}\quad,\quad \pi_b^i=\{k| k\in \pi^i, t_b^k\neq
  \phi, t_b^k < t_b^i\} 
\end{equation*}
$\mathcal{C}_b^i$ is the set of nodes that node $i$ has infected
before the end of the $b^{\text{th}}$ block, i.e., node $i$ is the
established parent of all nodes $j \in
\mathcal{C}_b^i$. $\pi_b^i\subseteq \pi^i$ is the
subset of potential parents of node $i$ who have been infected by the
end of the $b^{\text{th}}$ block. Table~\ref{ta:not} lists the
notation used in this paper.
The goal is to infer the set of infection parameters $\bx_b$ at the
 end of the $b^{\text{th}}$ block that best explains the received
 signals up to the end of block $b$, i.e., $\bd_{\bB_{1:b}}=\{\bd_{\bB_{1}}, ..., \bd_{\bB_{b}} \}$. More
 precisely, we aim to find the most probable set of parameters
 $(\hat{\bz}_b,\hat{\bt}_b,\hat{\balpha}_b)$ conditioned
 on the received signals $\bd_{\bB_{1:b}}$. We denote this set of
 Maximum A Posteriori (MAP) estimate of infection parameters by
 $\bx^{MAP}_b$: \begin{equation}\label{mainopt}
   \bx^{MAP}_b=(\hat{\bz}_b,\hat{\bt}_b,\hat{\balpha}_b)=
   \underset{(\bz_b,\bt_b,\balpha_b)}{\arg \max} \quad
   f(\bz_b,\bt_b,\balpha_b|\bd_{\bB_{1:b}}) \end{equation}

 Solving the optimization problem in \eqref{mainopt} is challenging,
 especially considering the fact that the data arrives in blocks and
 we need to make decisions before we have access to the entire
 signals. Hence, we resort to Monte Carlo Markov Chain (MCMC) methods
 to generate samples from the posterior distribution, $f(\bz_b,\bt_b,\balpha_b|\bd_{\bB_{1:b}})$, 
 and assess the underlying diffusion process based on these samples.
 We first describe the batch inference approach based on Gibbs Sampling (GS) that we proposed in
 \cite{shaghaghian2016bayesian}. We then extend this framework by
 considering the cases where no infection
 time is detected in the interval under study. We use this batch
 inference method to generate particles in the first received block of
 data. We then design an online inference algorithm based on
 Sequential Monte Carlo (SMC) techniques to find the set of infection
 parameters (i.e. network structure and cascade information) that best
 explains the observed data at the end of each block $b>1$. In order
 to do so, the particles for each block $b$ are obtained by updating
 particles of block $b-1$ using the received signals $\bd_{\bB_b}$. In
 order to have a unified notation throughout the paper, we
 use the $b$ batch sub-indices from now on for both the batch and
 sequential settings. For batch inference scenarios
 we have $\bx = \bx_1=( \bz_1,\bt_1,\balpha_1)$ since
 $N_B=1$ and $M=N_T$.

\section{Batch Inference Method}\label{batch}

Assuming that the entire data signal is available at each node, we
develop a batch (offline) inference algorithm based on Gibbs Sampling.
Using Bayes' rule and due to the dependencies we clarified in Section
\ref{model}, the joint conditional distribution
$f(\bx_1|\bd_{\bB_1})=f(\bz_1,\bt_1,\balpha_1|\bd_{\bB_1})$ is
\begin{equation}\label{Bayes}
  f(\bx_1|\bd_{\bB_1})=\frac{f(\bd_{\bB_1}|\bt_1,\bz_1,\balpha_1)f(\bt_1|\bz_1,\balpha_1)f(\bz_1|\balpha_1)f(\balpha_1)}{f(\bd_{\bB_1})}
\end{equation} In the rest of this section, we introduce appropriate prior
distributions for components of equation \eqref{Bayes}. The priors are
selected to allow flexibility in the incorporation of prior knowledge
and to facilitate computation. We then
use this Bayesian framework to develop methods for inferring the
underlying network structure as well as the cascade traces.

\subsection{Priors} \label{sec:prior}
As opposed to the previous methods and models (e.g.
\cite{gomez2011uncovering,embar2014bayesian,
  shaghaghian2016bayesian}), we accommodate scenarios
where an arbitrary node $i$ may never become infected over the study period of
length $N_T$. We indicate this case by assigning null values to the infection
time and parent of node $i$, i.e., $t_1^i=\phi$ and $z_1^i=\phi$. The
main building block of the diffusion model is the probability density
function $f(t_1^i|z_1^i,\alpha_1^{iz^i},t_1^{z^i})$, which models the
conditional likelihood that node $z_1^i$ (infected at time
$t_1^{z_1^i}$) transfers the infection to node $i$ at time
$t_1^i>t_1^{z_1^i}$. Intuitively, the ability of an infected node to
transfer the contagion to other nodes is expected to decay as time
passes. Hence, the monotonic memoryless exponential distribution is a
good candidate. An alternative way to motivate this model is to
consider repeated interactions between the nodes, each being a
Bernoulli trial with a small probability of infection. 

Some authors (e.g. \cite{gomez2011uncovering}) have studied the effect
of heavy tailed Power law or non-monotonic Rayleigh distributions on
the inference methods when the cascades are observed. For our prior
distribution, we employ the exponential decay assumption. Since
our observations $\bd$ are assumed to be discrete time series, we
choose the discrete counterpart of an exponential distribution for
$f(t_1^i|z_1^i,\alpha_1^{iz_1^i},t_1^{z_1^i})$, i.e., a geometric
distribution with parameter $p_1^i=1-e^{-\alpha_1^{iz_1^i}}$. 

Consider the case where $t_1^1 \geq t_1^2 \geq
\dots\geq t_1^N$, employing the convention that the null infection
time $\phi>N_T$. We have $f(\bt_1|\bz_1,\balpha_1)=\prod_{i \in
  \cN} f(t_1^i|z_1^i,\balpha_1,t_1^{i+1:N})$ where
  \begin{equation}\label{prior_t}
  f(t_1^i|z_1^i,\balpha_1,t_1^{i+1:N})=\begin{cases} 
 p_1^i (1-p_1^i)^{t_1^i-t_1^{z_1^i}-1}   & 1 \leq t_1^i \leq N_T,\\
(1-p_1^i)^{N_T} & \quad t_1^i =\phi.
\end{cases}
\end{equation}
The expression for $t_1^i =\phi$ is simply one minus the value of the cumulative
distribution function of the random variable $t_1^i$ at time $N_T$.
A similar expression can be obtained for any arbitrary ordering of the changepoints.

We assume that, conditioned on the link strengths, the probability
that one of the nodes in the candidate parent set $v \in \pi_i$ is the
actual parent $z_i$ of node $i$ is independent of the probability that
a node $r \in \pi_j$ is the parent $z_j$ of node $j$. The exponential decay assumption and the fact that
$z_1^i$ is the first node that transfers the infection to node $i$
makes the multinomial distribution a good candidate for capturing the prior
distribution of parents given the link strength values:
\begin{equation}\label{prior_z} f(\bz_1|\balpha_1)=\prod_{i \in \cN}
  f(z_1^i|\alpha_1^{ij},{j \in \pi^i})=\prod_{i \in \cN}
  \frac{\alpha_1^{iz_1^i}}{\sum_{j\in\pi^i}\alpha_1^{ij}}
\end{equation}

As for the prior distribution for strength value of link $ij$, we
choose a Gamma distribution, i.e., $\alpha_1^{ij} \sim
\Gamma(\kappa_{ij},\theta_{ij})$. This choice of prior
distribution allows us to model a wide range of $\alpha$ values for
different links of the network. We can capture both highly informed
knowledge about strong links or the uninformed case where we have
little prior information about the strength of links. Therefore, assuming that link strength
values of different links are independent, we have
\begin{equation}\label{prior_alpha} f(\balpha_1)=\prod_{i\in \cN , j
    \in \pi^i} f(\alpha_1^{ij})= \prod_{i\in \cN , j \in \pi^i}
  \frac{x^{\kappa_{ij}-1}e^{-\frac{x}{\theta_{ij}}}}{\Gamma(\kappa_{ij}){\theta_{ij}}^{\kappa_{ij}}}
\end{equation}

Finally, we assume that node $i$'s observed data, $\bd_{\bB_1}^i$, are
conditionally independent of the observations from all other nodes and that they
follow the same prior distribution with two different sets of
hyperparameters $\bgam_1=(\gamma_1^1,\dots,\gamma_1^N)$,
$\bgam_2=(\gamma_2^1,\dots,\gamma_2^N)$ before and after being
infected at $t^i$. Hence,
\begin{equation}\label{prior_d}
\begin{aligned}
f(\bd_{\bB_1}|\bz_1,\bt_1,\balpha_1)&=\prod_{i \in \cN}f(\bd_{\bB_1}^i|t_1^i)\\
&=\prod_{i \in \cN}f(d^i_{1:t^i};\gamma_1^i)f(d^i_{t^{i+1}:N_T};\gamma_2^i)
\end{aligned}
\end{equation}
The choice of prior $f(d^i_{1:t^i};\gamma_1^i)$ depends on the
application. We test cases where the data are independently drawn from Gaussian and Poisson distributions in our numerical
simulations in Section \ref{res}, but more complex structure can be
readily incorporated in the time-series model (autoregressive
processes, for example). In the next section, we use the
prior distributions described above to design batch and online
inference methods in a Bayesian framework.

 \subsection{Gibbs Sampling}

 With the proposed distributions in \eqref{prior_t}-\eqref{prior_d},
 we can calculate the probability of any arbitrary set $( \bz_1,
 \bt_1, \balpha_1)$ up to a constant $\frac{1}{f(\bd_{\bB_1})}$
 using \eqref{Bayes}. We use Gibbs Sampling (GS) to generate $N_s$
 samples from the posterior distribution of \eqref{Bayes}. In other
 words, we use full conditional distributions for each of
 the infection parameters $z_1^i,t_1^i,\alpha_1^{ij}$ ($i, j \in \cN$)
 to generate samples. We denote the parents and infection times of all
 the nodes in the network except node $i$ respectively by
 $\bz_1^{\overline{i}}$, $\bt_1^{\overline{i}}$. Also, the vector of
 link strengths of all node $i$'s links except the link between nodes
 $i$ and $j$ is denoted by $\balpha_1^{\overline{ij}}$ . The full
 conditional probabilities for GS are as follows.

{ \bf a)} For the parent of node $i \in \cN$
\begin{equation}
f(z_1^i|\bd_{\bB_1},\bz_1^{\overline{i}},\bt_1,\balpha_1) \propto f(t_1^i|z_1^i,\alpha_1^{iz_1^i},t_1^{z_1^i}) f(z_1^i|\alpha_1^{ij})_{j \in \pi^i}
\end{equation}

{ \bf b)} For the infection time of node $i \in \cN$
\begin{equation}
\begin{aligned}
f(t_1^i|\bd_{\bB_1},\bz_1,&\bt_1^{\overline{i}},\balpha_1) \propto \\
&f(\bd_{\bB_1}^i|t_1^i) f(t_1^i|z_1^i,\alpha_1^{iz_1^i},t_1^{z_1^i})\prod_{k \in \cC_1^i} f(t_1^k|\alpha_1^{ki},t_1^i)
\end{aligned}
\end{equation}

{ \bf c)} For the link strength between nodes $i\in \cN$ and $j \in \pi^i$
\begin{equation}
f(\alpha_1^{ij}|\bd_{\bB_1},\bz_1,\bt_1,\balpha^{\overline{ij}}_1) \propto f(t_1^i|z_1^i,\alpha_1^{iz_1^i},t_1^{z_1^i})f(z_1^i|\balpha_1) f(\alpha_1^{ij})
\end{equation}
Building upon this batch inference method, we propose an online
inference method in the next section. In this online approach, the
batch inference method is used to make decisions about diffusion
parameters in the first received block of data.

\section{Online Inference Method}\label{online}
In the online setting, the goal is to compute the
filtering posterior distribution $f(\bx_b|\bd_{\bB_{1:b}})$. In the Bayesian framework we have,
\begin{equation}\label{integral_first}
f(\bx_b|\bd_{\bB_{1:b}}) \propto \int f(\bd_{\bB_b}|\bx_b)f(\bx_b|\bx_{b-1})f(\bx_{b-1}|\bd_{\bB_{1:b-1}})d_{\bx_{b-1}}
\end{equation}
Since calculating \eqref{integral_first} is analytically intractable in our
application, we use the sequential Markov Chain Monte Carlo (SMCMC) framework proposed in
\cite{septier2009mcmc} to obtain an approximation of this filtering
distribution. 
In this framework, instead of striving to directly sample from
$f(\bx_b|\bd_{\bB_{1:b}})$, which has complexity issues due to the
need to perform the marginalization captured by the integral
in~\eqref{integral_first}, we sample from the joint posterior distribution
$f(\bx_b,\bx_{b-1}|\bd_{\bB_{1:b}})$ where
\begin{equation}\label{app2_1}
f(\bx_b,\bx_{b-1}|\bd_{\bB_{1:b}})=\frac{f(\bd_{\bB_b}|\bx_b)f(\bx_b|\bx_{b-1})f(\bx_{b-1}|\bd_{\bB_{1:b-1}})}{f(\bd_{\bB_b}|\bd_{\bB_{1:b-1}})}
\end{equation}

In the sequential MCMC approach, we maintain a set of particles that
provides a sample-based approximation to the distribution of interest
$f(\bx_{b}|\bd_{\bB_{1:b}})$. After processing block {$b-1$}, since
the posterior distribution $f(\bx_{b-1}|\bd_{\bB_{1:b-1}})$ does not
have a closed form representation, it is approximated with an
empirical distribution based on the particle set $\{\bx_{b-1}^j
|j=1,\dots,N_s\}$:
 \begin{equation}
f(\bx_{b-1}|\bd_{\bB_{1:b-1}})=\frac{1}{N_s}\sum_{j=1}^{N_s}\delta(\bx_{b-1}-\bx_{b-1}^{j})
\end{equation}
Hence, the target distribution $f(\bx_b,\bx_{b-1}|\bd_{\bB_{1:b}})$
that we wish to sample from can be approximated as 
\begin{equation}\label{target}
\begin{aligned}
\begin{cases}
\frac{f(\bd_{\bB_b}|\bx_b)f(\bx_b|\bx_{b-1}^{s})}{\sum_{j=1}^{N_s}f(\bd_{\bB_b}|\bx_{b-1}^{j})}
& \text{if } \bx_{b-1} \in \{\bx_{b-1}^{s}|s=1,\dots,N_s\}  ,\\
0 &  \text{otherwise}.
\end{cases}
\end{aligned}
\end{equation}
Here, $f(\bx_b|\bx_{b-1}^{s})$ is referred to as the transition
distribution. We will derive an appropriate expression for this probability
distribution in Section~\ref{sectransition}.

Algorithm \ref{algo:online} presents our proposed online inference
method based on the MCMC-based particle algorithm used in
\cite{septier2009mcmc}. As mentioned earlier, we use the batch
inference method described in section \ref{batch} to generate a set of
$N_s$ particles in the first block of the data and use it as the
initial particle set for the SMC procedure of next blocks. In each
block $b>1$ of the data, the SMC procedure consists of two main steps.
The first step, {\it joint draw}, is a joint Metropolis-Hastings (MH)
proposal step in which instead of sampling from
$f(\bx_b,\bx_{b-1}|\bd_{\bB_{1:b}})$, we sample from the
proposal distribution $q(\bx_b,\bx_{b-1}|\bx_b^{m-1},\bx_{b-1}^{m-1})$
and accept the proposed samples with probability $\rho$. Lines 11-13 of
Algorithm \ref{algo:online} describe the components of the proposal
distributions. The proposal distributions and the MH acceptance ratio
are respectively derived in \ref{proposal} and \ref{MH}. The second
step of the SMC procedure, {\it refinement}, is an individual
refinement GS step where $\bx_b$ is updated by sampling from
$f(\bx_b|\bx_{b-1}^{m},\bd_{\bB_b})$.

\begin{algorithm}[!htb]
\caption{ SMC-Based Online Inference Method} \label{algo:online}
\begin{spacing}{1.0}
\begin{algorithmic}[1]
\vspace{0mm}
\STATE \texttt{// First block}
\STATE Generate $N_s$ samples $\bx_1^{s}|_{s=1,\dots,N_s} \sim f(\bx_1|\bd_{\bB_1})$ using GS
\STATE \texttt{// Next blocks}
\STATE Initialize the particle set $\cP=\{\bx_1^{s}|s=1,\dots,N_s\}$
\STATE Set $N_{MCMC}=N_{thin}\times N_s+N_{burn}$
\FOR{\texttt{$b=2,\dots,N_b$}}
\STATE Calculate $\bt_b^{ML}$ using \eqref{t_ML}
\FOR{\texttt{$m=1,\dots,N_{MCMC}$}}
\STATE \texttt{// Joint Draw}
\STATE Propose $\bx_{b-1}^*$ uniformly from $\cP$
\STATE Propose $\bt_b^* \sim h_{\bt}(\bt_b|\bx_{b-1}^*;\bt_b^{ML})$ using \eqref{h_t}
\STATE Propose $\balpha_b^* \sim h_{\balpha}(\balpha_b|\bx_{b-1}^*,\bt_b^*)$ using \eqref{h_alpha}
\STATE Propose $\bz_b^* \sim h_{\bz}(\bz_b|\bx_{b-1}^*,\bt_b^*,\balpha_b^*)$ using \eqref{h_z}
\STATE Set $\bx_b^*=( \bz_b^*,\bt_b^*,\balpha_b^* )$
\STATE Calculate MH acceptance probability $\rho$ using \eqref{rho}
\STATE Accept $(\bx_b^{m},\bx_{b-1}^{m})=(\bx_b^*,\bx_{b-1}^*)$ with probability $\rho$
\STATE \texttt{// Refinement}
\STATE Generate $\bx_b^* \sim f(\bx_b|\bx_{b-1}^{m},\bd_{\bB_b})$ using GS and set $\bx_b^m=\bx_b^*$
\ENDFOR
\STATE Set $\cP=\{\bx_b^{N_{burn}+kN_{thin}}|k=1, \dots,N_s\}$
\STATE Set $\bx_b^{MAP}$ as the most repeated particle in $\cP$
\ENDFOR
\end{algorithmic}
\end{spacing}
\end{algorithm}

Finally, { \it thinning} and {\it burn-in} procedures are performed by
storing one out of every $N_{thin}$ accepted samples after discarding
the initial $N_{burn}$ samples. In the rest of this section, we derive
the transition and proposal distributions and use them to calculate
the MH acceptance ratio. We then explain the details of the refinement
step and conclude the section by commenting on the computational
complexity of Algorithm \ref{algo:online}.
\vspace{-0.4cm}
\subsection{Transition Distribution} 
\label{sectransition}
Using the framework explained in Section \ref{model}, we have
\begin{equation}\label{transition}
\begin{aligned}
f(\bx_b|\bx_{b-1}^{s})&=f(\bz_b,\bt_b|\bx_{b-1}^s,\balpha_b)f(\balpha_b|\bx_{b-1}^s)\\
&=\prod_{i \in \cN} f(t_b^i,z_b^i|\bx_{b-1}^s,\balpha_b)\prod_{k\in\pi^i} f(\alpha_b^{ik})
\end{aligned}
\end{equation}
Due to the characteristics of the online method,
once a non-null value has been assigned to the infection parameters of
a node in block $b$, this decision is respected in the following
blocks. More precisely,
\begin{equation}\label{transition1}
f(\alpha_b^{ik})=\begin{cases}
\delta(\alpha_b^{ik}-\alpha_{b-1}^{ik})&\text{if }  t_{b-1}^i \neq \phi,\\
\Gamma(\kappa_{ik},\theta_{ik})&\text{if }  t_{b-1}^i = \phi,
\end{cases}
\end{equation}
and
\begin{equation}\label{transition2}
\begin{aligned}
&f(t_b^i,z_b^i|\bx_{b-1}^s,\alpha_b^{ij})_{j\in \pi_b^i}=\\
&\begin{cases}
\delta(t_b^i-t_{b-1}^i)\delta(z_b^i-z_{b-1}^i) & \text{if }  t_{b-1}^i
\neq \phi, \vspace{0.2cm}\\
\begin{tabular}{@{}c@{}}$ \sum_{l \in \pi_b^i}\sum_{x =t_b^l+1}^{\bB_b[M]} g_1(l,x,i)+\delta(z_b^i-\phi)\times$\\
$\delta(t_b^i-\phi)(1-\sum_{l \in \pi_b^i}\sum_{x =t_b^l+1}^{\bB_b[M]}g_1(l,x,i))$\end{tabular} &\text{if }  t_{b-1}^i = \phi,
\end{cases}
\end{aligned}
\end{equation}
where 
\begin{equation}\label{transition3}
g_1(l,x,i)=\frac{\alpha_b^{il}}{\sum_{k \in \pi^i}\alpha_b^{ik}}(p_b^i)(1-p_b^i)^{x-t_b^l-1}\delta(z_b^i-l)\delta(t_b^i-x)
\end{equation}
and $p_b^i=1-e^{-\alpha_b^{iz_b^i}}$. According to
\eqref{transition2}, if node $i$ has not become infected by the end of
block $b-1$, the probability that it becomes infected by node $l \in \pi_b^i$ at some time $x
\in [t_b^l, \bB_b[M]]$ is $g_{1}(l,x,i)$. Equation
\eqref{transition3} shows that this probability equals the product of
the geometric and multinomial distributions respectively defined in
\eqref{prior_t} and \eqref{prior_z}. Node $i$ remains susceptible
(i.e., $t_{b}^i=\phi$) with a probability that equals one minus the
sum of the probabilities of being infected by each node $l \in \pi_i$
at each time step $x \in [t_b^l, \bB_b[M]]$.

\subsection{Proposal Distribution}\label{proposal}
As presented in line 9 of Algorithm \ref{algo:online}, the optimal
importance sampling distribution (assuming that we adopt the
particle-based approximation for $f(\bx_{b-1}|\bd_{\bB_{1:b-1}})$) is:
\begin{equation}
\begin{aligned}
&q(\bx_b,\bx_{b-1}|\bx_b^{m-1},\bx_{b-1}^{m-1})=f(\bx_b|\bx_{b-1},\bd_{\bB_b})f(\bx_{b-1}|\bd_{\bB_{1:b-1}})\\
&=\begin{cases}
\frac{f(\bd_{\bB_b}|\bx_b)f(\bx_b|\bx_{b-1})}{f(\bd_{\bB_b}|\bx_{b-1})} & \text{if } \bx_{b-1} \in \{\bx_{b-1}^s,s=1,\dots,N_s\}, \label{proposal1A}\\
0 &  \text{otherwise}.
\end{cases}
\end{aligned}
\end{equation}
Although calculating the predictive density in the denominator
of~\eqref{proposal1A} is not necessary for sampling, it is eventually
required for calculating the acceptance ratio. To avoid numerical
integration of the predictive density at every iteration, we benefit
from an auxiliary parameter which can be obtained from the data. In
each block $b$, we calculate the maximum likelihood changepoint of
each individual time series $\bd_{\bB_b[1]:\bB_b[M]}$ for all $i \in
\cN$. We denote the vector of these maximum likelihood changepoints by
$\bt_b^{ML}=[t_b^{ML^1},\dots,t_b^{ML^N}]^T$, where
\begin{equation}\label{t_ML}
t_b^{ML^i} \triangleq \arg \underset{\bB_b[1]\leq t \leq \bB_b[M]}{\max} f(\bd_{\bB_b[1]:t};\gamma_1^i)f(\bd_{t+1:\bB_b[M]};\gamma_2^i)
\end{equation}
Using this auxiliary parameter, we design the following proposal distribution.
\begin{equation}\label{proposal2}
\begin{aligned}
&q(\bx_b,\bx_{b-1}|\bx_b^{m-1},\bx_{b-1}^{m-1})=h(\bx_b|\bx_{b-1};\bt_b^{ML})f(\bx_{b-1}|\bd_{\bB_{1:b-1}})\\
&=\begin{cases}
\frac{1}{N_s}h(\bx_b|\bx_{b-1};\bt_b^{ML}) & \text{if } \bx_{b-1} \in \{\bx_{b-1}^s,s=1,\dots,N_s\}, \\
0 &  \text{otherwise}.
\end{cases}
\end{aligned}
\end{equation}
where 
\begin{equation}\label{h}
\begin{aligned}
&h(\bx_b|\bx_{b-1};\bt_b^{ML})=\\
&h_{\bt}(\bt_b|\bx_{b-1};\bt_{b}^{ML})h_{\balpha}(\balpha_b|\bt_b,\bx_{b-1})h_{\bz}(\bz_b|\bt_b,\balpha_b,\bx_{b-1}).
\end{aligned}
\end{equation}
Each component of $h(\bx_b|\bx_{b-1};\bt_b^{ML})$ is the proposal distribution for one of the infection parameters. For each node $i \in \cN$, we independently propose an infection time whose distance from the maximum likelihood $t_b^{ML^i}$ follows a geometric distribution with pre-defined parameter $r_b^i$. In other words, we sample from $h_{\bt}(\bt_b|\bx_{b-1};\bt_{b}^{ML})=\prod_{i \in \cN} h_{t}^i(t_b^i|\bx_{b-1};t_{b}^{ML^i})$, where
\begin{equation}\label{h_t}
\begin{aligned}
&h_{t}^i(t_b^i|\bx_{b-1};t_{b}^{ML^i})=\\
&\begin{cases}
\delta(t_b^i-t_{b-1}^i) & t_{b-1}^i \neq \phi, \\
\sum_{n=1}^M g_2(n,i)+(1-\sum_{n=1}^Mg_2(n,i))\delta(t_b^i-\phi) & t_{b-1}^i = \phi.
\end{cases}
\end{aligned}
\end{equation}
and
\begin{equation}
g_2(n,i)=\frac{1}{2}r_b^i(1-r_b^i)^{|\bB_b[n]-t_b^{ML^i}|} \delta(t_b^i-\bB_b[n]).
\end{equation}
Using the proposed $\bt_b^*$, we propose $\balpha_b^*$ by sampling from $h_{\balpha}(\balpha_b|\bt_b^*,\bx_{b-1})=\prod_{i \in \cN} \prod_{k \in \pi^i} h_{\alpha}^i(\alpha_b^{ik}|\bt_b^*,\bx_{b-1})$ where
\begin{equation}\label{h_alpha}
h_{\alpha}^i(\alpha_b^{ik}|\bt_b^*,\bx_{b-1})=\begin{cases}
\delta(\alpha_b^{ik}-\alpha_{b-1}^{ik}) &  \text{if } t_{b}^i \neq \phi , t_{b-1}^i \neq \phi,\\
\Gamma(\kappa_{ki},\theta_{ki}) &  \text{if } t_{b-1}^i = \phi.
\end{cases}
\end{equation}
Finally, we propose $\bz_b^*$ by generating samples from $h_{\bz}(\bz_b|\bt_b^*,\balpha_b^*,\bx_{b-1})=\prod_{i \in \cN}  h_{z}^i(z_b^{i}|\bt_b^*,\balpha_b^*,\bx_{b-1})$,  i.e.,
\begin{equation}\label{h_z}
\begin{aligned}
&h_{z}^i(z_b^i|\bt_b^*,\balpha_b^*,\bx_{b-1})=\\
&\begin{cases}
\delta(z_b^{i}-z_{b-1}^{i}) &  \text{if } t_{b}^i \neq \phi , t_{b-1}^i \neq \phi,\\
\delta(z_b^{i}-\phi) &  \text{if } t_{b}^i = \phi,t_{b-1}^i = \phi,\\
\begin{tabular}{@{}c@{}}$ \sum_{l \in \pi_b^i}\frac{\alpha_b^{il}}{\sum_{j \in \pi^i}\alpha_b^{ij}}\delta(z_b^i-l)$+ \\ $(1-\frac{\sum_{l \in \pi_b^i}\alpha_b^{il}}{\sum_{j \in \pi^i}\alpha_b^{ij}})\delta(z_b^i-\phi)$\end{tabular} &  \text{if } t_{b}^i \neq \phi,t_{b-1}^i = \phi.
\end{cases}
\end{aligned}
\end{equation}
Proposition \ref{prop1} shows that the proposal $q$ is normalized. The proof is provided in Appendix \ref{pp1}.
\begin{proposition}\label{prop1}
$h(\bx_b|\bx_{b-1};\bt_b^{ML})$ is a probability distribution
function. In particular,  
\begin{equation}\label{integral}
\int \sum_{\bt_b \in \cS_1\times \dots\times \cS_1}\sum_{\bz_b \in \cS_2\times \dots\times \cS_2} h(\bx_b|\bx_{b-1};\bt_b^{ML}) d_{\balpha_b}=1
\end{equation}
where $\cS_1=\{\bB_1[1],\dots,\bB_b[M],\phi\}$ and $\cS_2=\pi_b^i\cup\{\phi\}$.
\end{proposition}

\subsection{Metropolis-Hastings Acceptance Ratio}\label{MH}
The acceptance ratio of the Metropolis-Hastings sampling approach proposed in \cite{hastings1970monte} is \begin{equation}\label{rho_init}
\begin{aligned}
&\rho=\\
&\min\left(1,\frac{f(\bx_b^*,\bx_{b-1}^*|\bd_{\bB_{1:b}})}{q(\bx_b^*,\bx_{b-1}^*|\bx_b^{m-1},\bx_{b-1}^{m-1})}\frac{q(\bx_b^{m-1},\bx_{b-1}^{m-1}|\bx_b^*,\bx_{b-1}^*)}{f(\bx_b^{m-1},\bx_{b-1}^{m-1}|\bd_{\bB_{1:b}})}\right)
\end{aligned}
\end{equation}
Replacing \eqref{target} and \eqref{proposal2} in the second argument of \eqref{rho_init}, we have
\begin{equation}\label{rho}
\frac{f(\bd_{\bB_b}|\bx_b^*)}{f(\bd_{\bB_b}|\bx_b^{m-1})}\frac{f(\bx_b^*|\bx_{b-1}^*)}{f(\bx_b^{m-1}|\bx_{b-1}^{m-1})}\frac{h(\bx_b^{m-1}|\bx_{b-1}^{m-1};\bt_b^{ML})}{h(\bx_b^*|\bx_{b-1}^*;\bt_b^{ML})}
\end{equation}
In evaluating this expression, the first ratio can be calculated
directly from the likelihood expressions. The second ratio can be
evaluated from \eqref{transition}, using
\eqref{transition1}-\eqref{transition3}. The third ratio can be
calculated from \eqref{h}, using \eqref{h_t}-\eqref{h_z}.

\subsection{Refinement Step}\label{Refine}
The Gibbs sampling procedure in the refinement step is basically the same as the
Gibbs sampling used in the batch inference. For each node $i \in \cN$, each
one of the diffusion parameters ($t_b^i,z_b^i,\alpha_b^i$) is treated
as a separate parameter block and is sampled using full conditional
distributions. However, some of the online inference features need to
be accounted for while deriving the full conditional distributions. These
distributions are derived in Appendix \ref{FCDR}.

\subsection{Computational Complexity}\label{Comple}
Line 7 of Algorithm \ref{algo:online} only needs to be performed once per
block. In each block $b$, we need to evaluate the probability
distributions of \eqref{t_ML} for each node at
each time $\bB_b[1]\leq t \leq \bB_b[M]$. Therefore, the computational
complexity of finding $\bt_b^{ML}$ for all blocks is of order $O(N_bNM)$. Lines 11 to
13 of the algorithm involve generating respectively $N{-}1$ geometric,
$\sum_{i \in \cN} |\pi_i|$ gamma, and $N{-}1$ multinomial random
variables in the worst case. These random number generation procedures
also exist in the refinement step (Line 18) and the Gibbs Sampling of
the first block (Line 2). In practice, we observe that generating a gamma
distributed random variable is significantly more computationally
expensive than generating the
other two random variables. Therefore, if we bound $\sum_{i \in \cN}
|\pi_i|$ by $N^2$, we observe that, depending on the values of $M$,
$N$, and $N_{MCMC}$, Algorithm \ref{algo:online} has computational
complexity of $O(N_b N_{MCMC} N^2)$ or $O(N_bNM)$. The computation
grows linearly with the number of MCMC iterations and is proportional
the square of the number of nodes in the network.

\section{Experiments}\label{res}
In this section, we investigate the efficiency of our proposed approach in modelling and solving the inference problem
in different diffusion network scenarios. We first test
our method in synthetic networks where the modelling assumptions are
exactly met and we know the ground truth. Then, we use the approach to
analyze two real-world datasets.

\begin{figure*}[t]
\centering
\subfloat[{\em A}: $\mu_2=100$, $\kappa_2=40$]{\includegraphics[width=.21\textwidth]{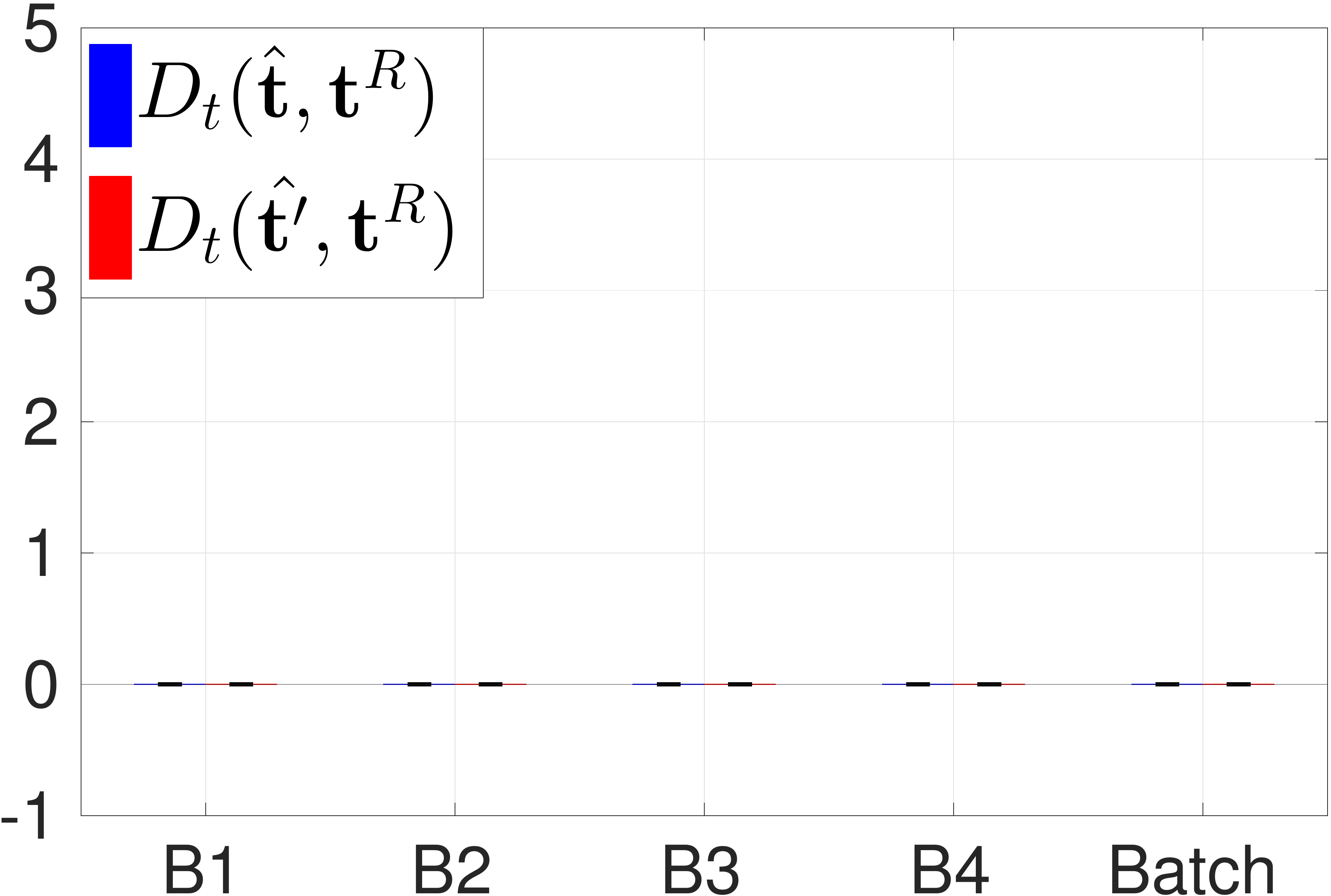}\label{fig:fig_t1}}\hspace*{0.17in}
\subfloat[{\em B}: $\mu_2=100$, $\kappa_2=2$]{\includegraphics[width=.21\textwidth]{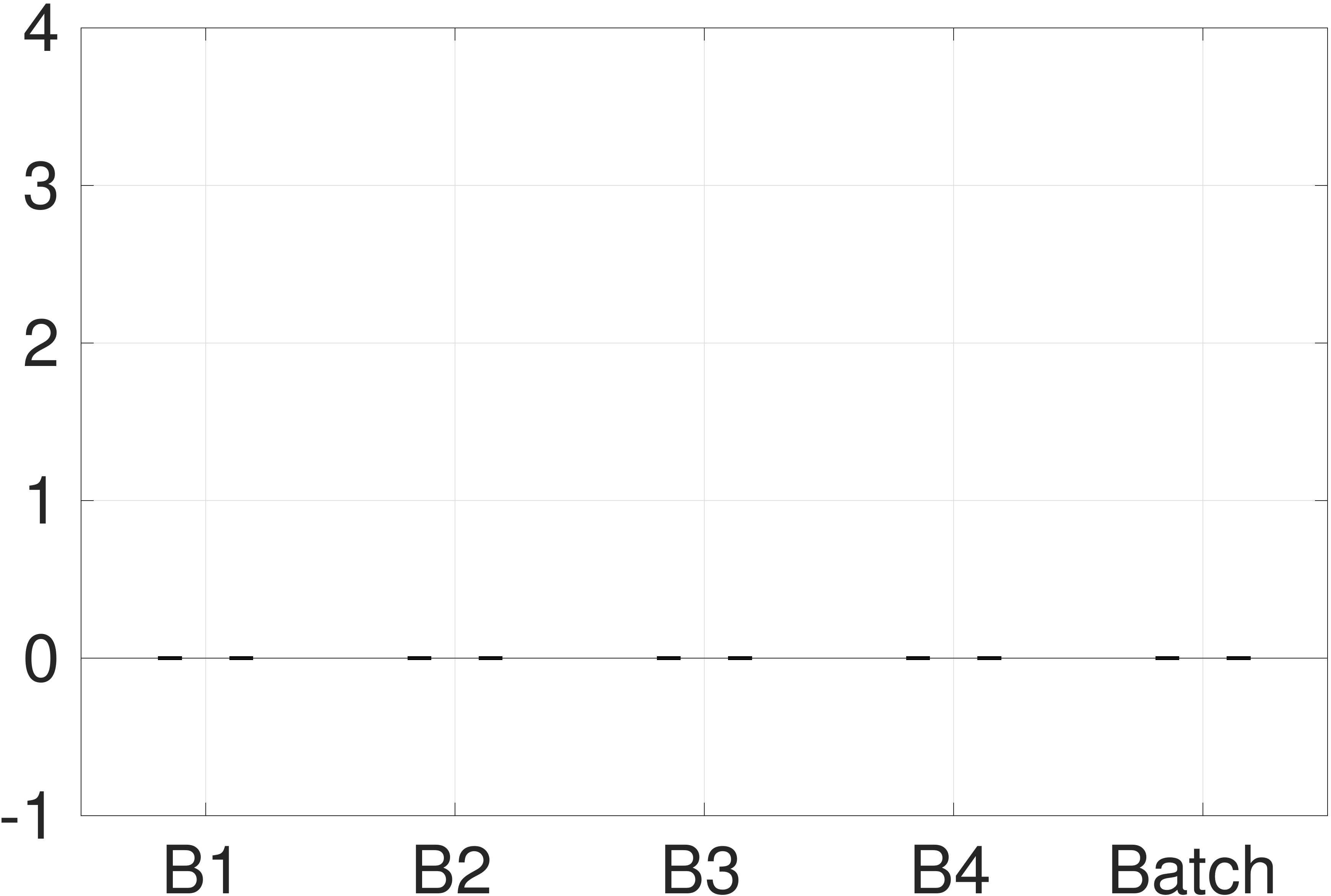}\label{fig:fig_t2}}\hspace*{0.17in} \subfloat[{\em C}: $\mu_2=11$, $\kappa_2=40$]{\includegraphics[width=.21\textwidth]{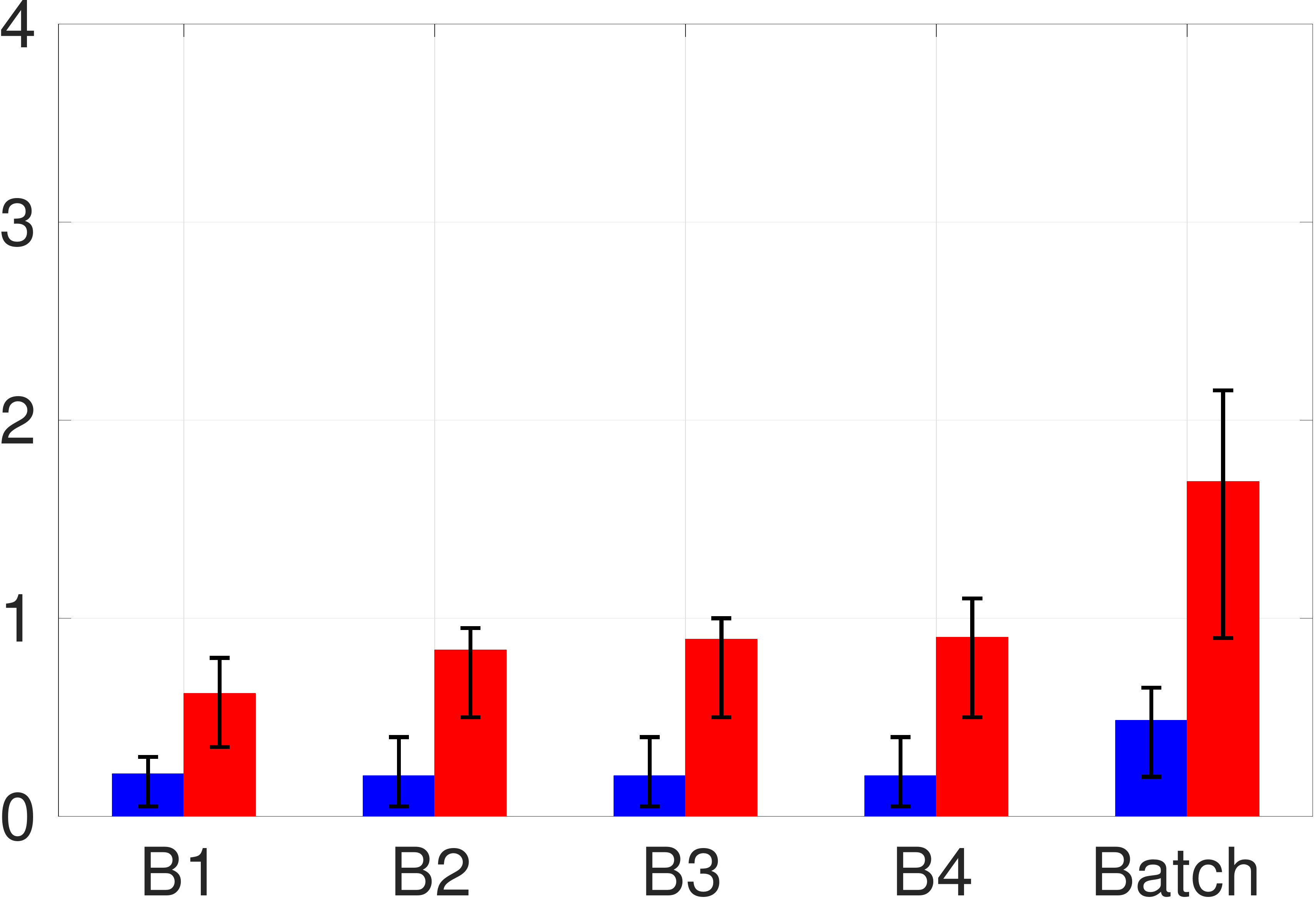}\label{fig:fig_t3}}\hspace*{0.17in}
\subfloat[{\em D}: $\mu_2=11$, $\kappa_2=2$]{\includegraphics[width=.21\textwidth]{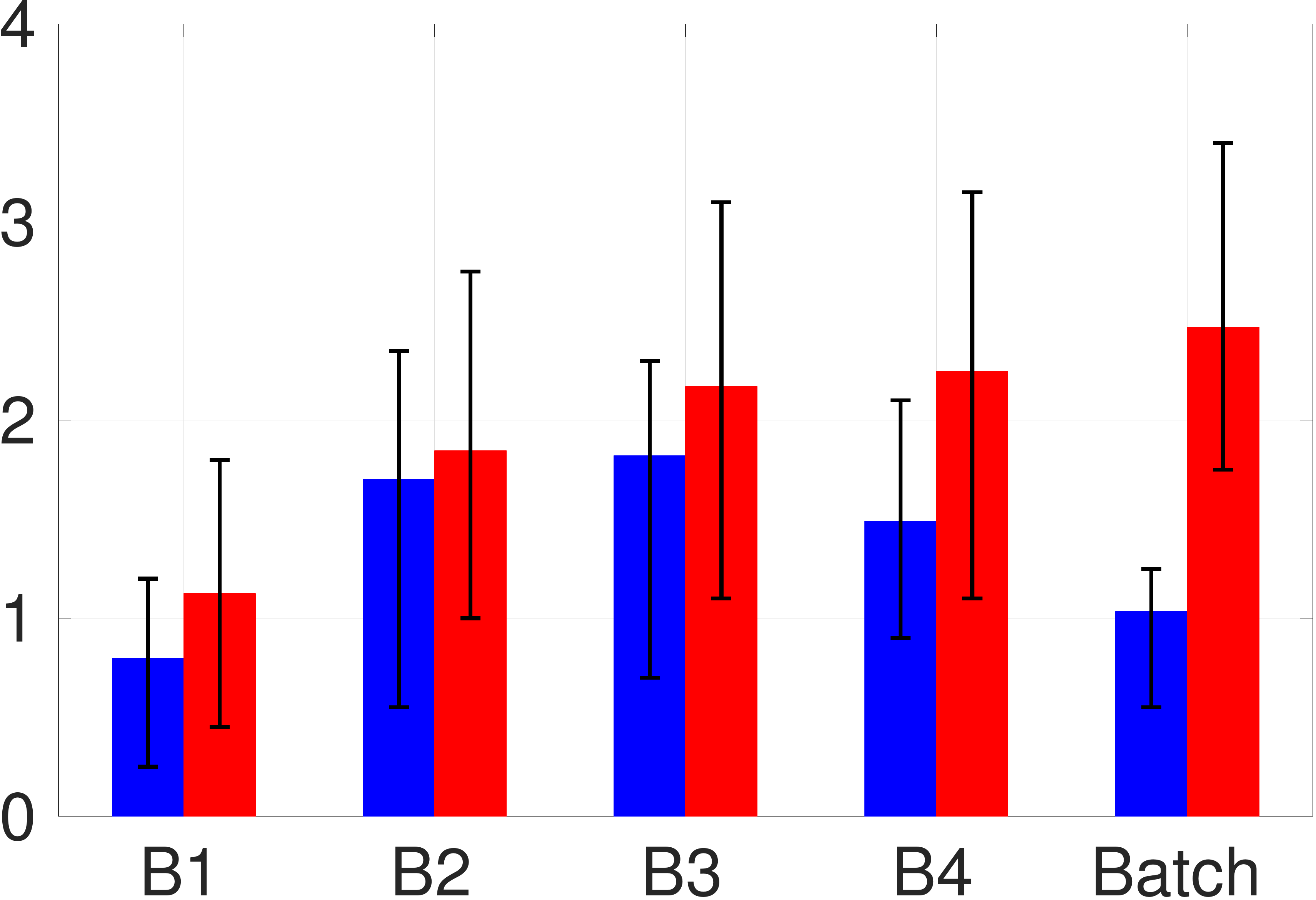}\label{fig:fig_t4}}
\caption{Deviations in detection of infection times}
\label{fig:t}
\end{figure*}
\begin{figure*}[t]
\centering
\subfloat[{\em A}: $\mu_2=100$, $\kappa_2=40$]{\includegraphics[width=.21\textwidth]{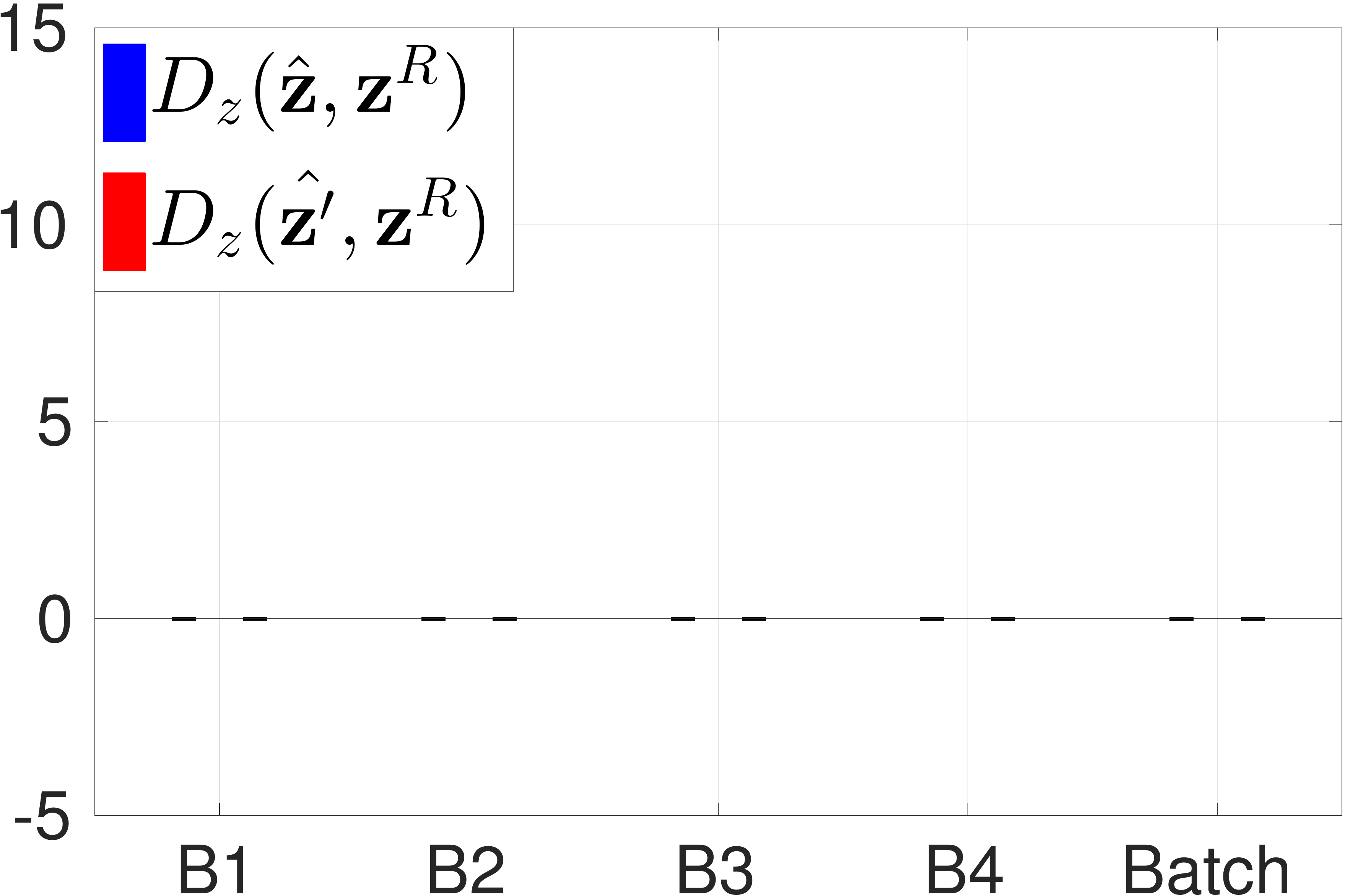}\label{fig:fig1_first_case}}\hspace*{0.17in}
\subfloat[{\em B}: $\mu_2=100$, $\kappa_2=2$]{\includegraphics[width=.21\textwidth]{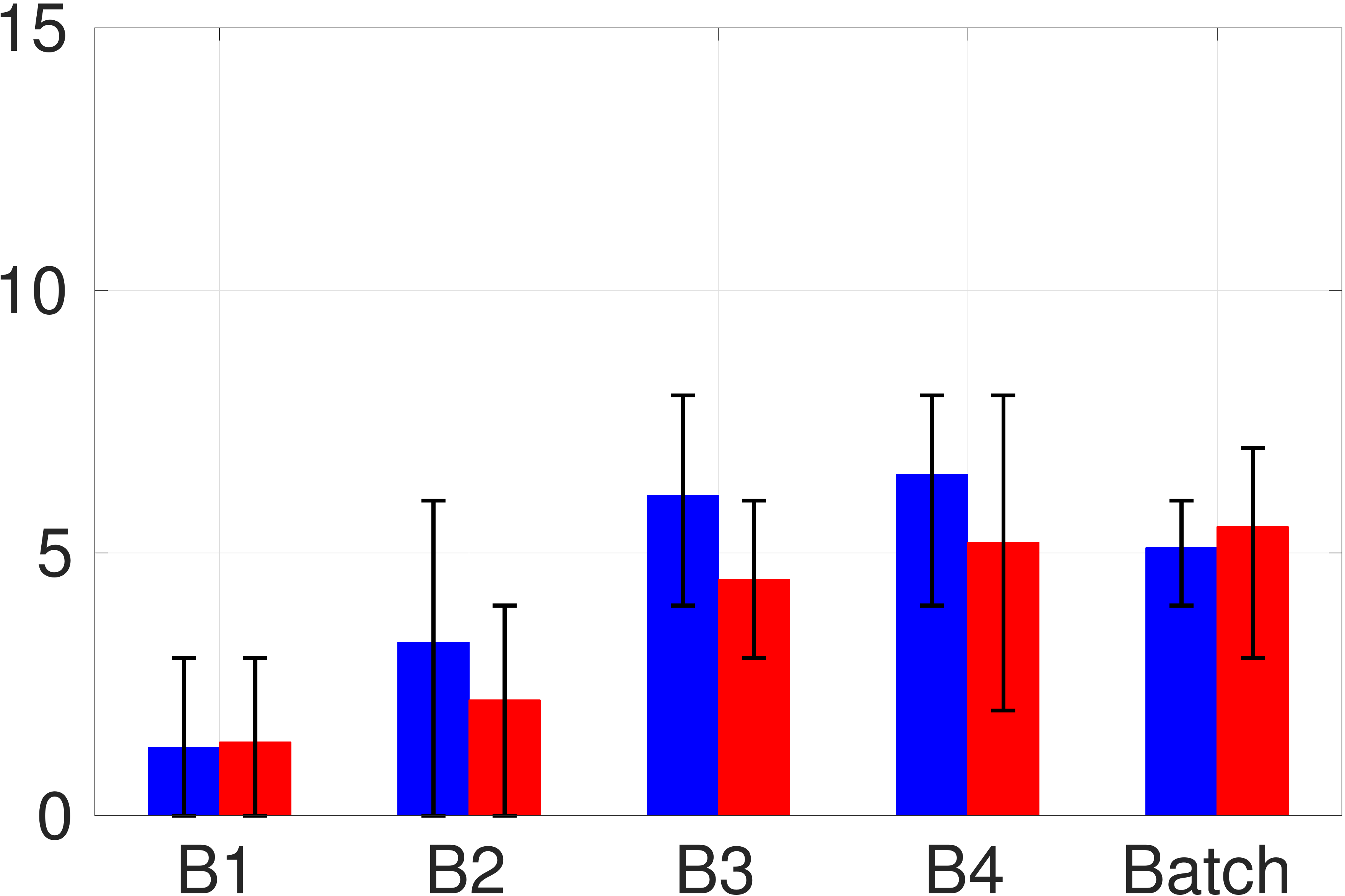}\label{fig:fig1_second_case}}\hspace*{0.17in} \subfloat[{\em C}: $\mu_2=11$, $\kappa_2=40$]{\includegraphics[width=.21\textwidth]{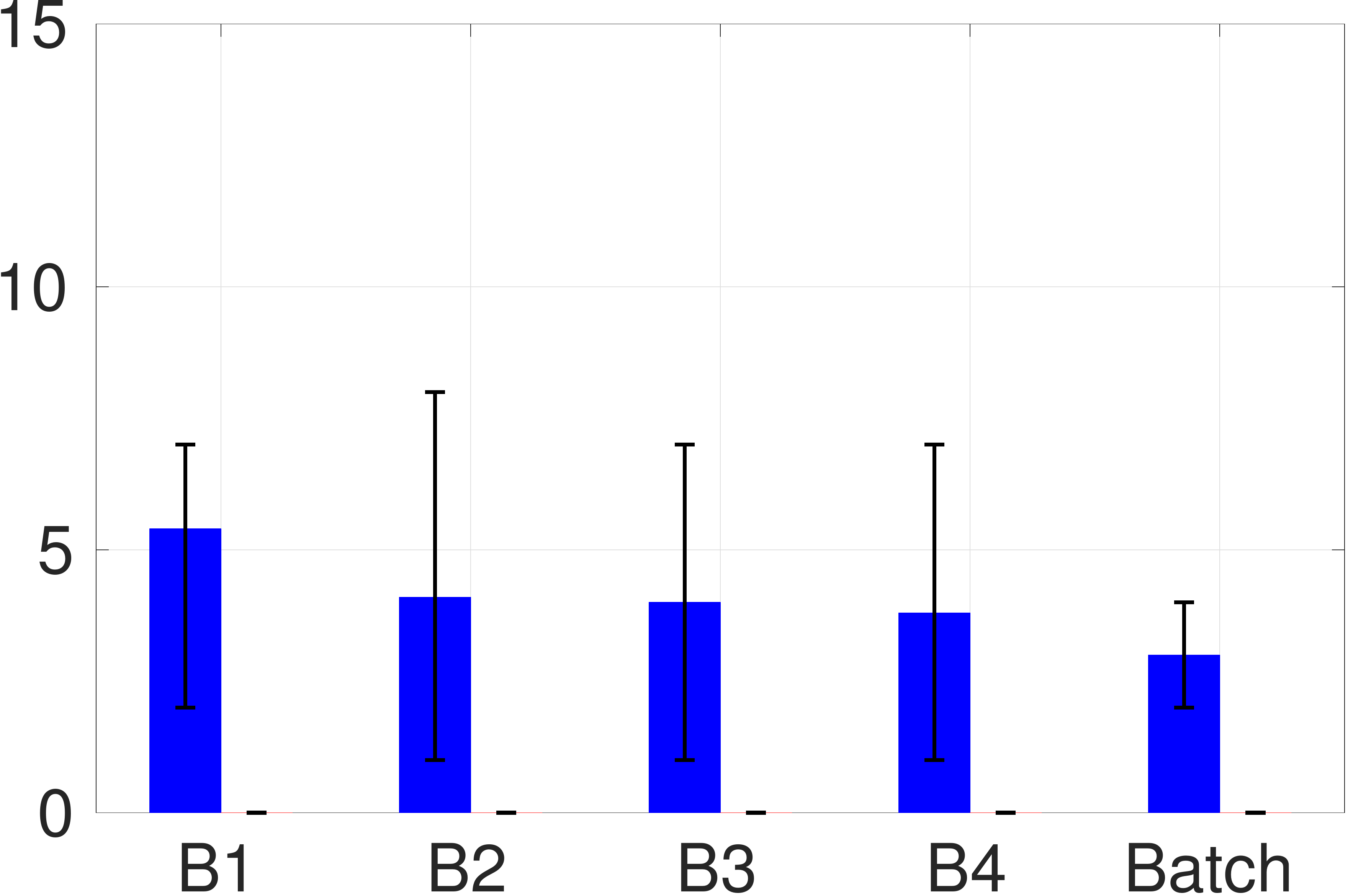}\label{fig:fig1_third_case}}\hspace*{0.17in}
\subfloat[{\em D}: $\mu_2=11$, $\kappa_2=2$]{\includegraphics[width=.21\textwidth]{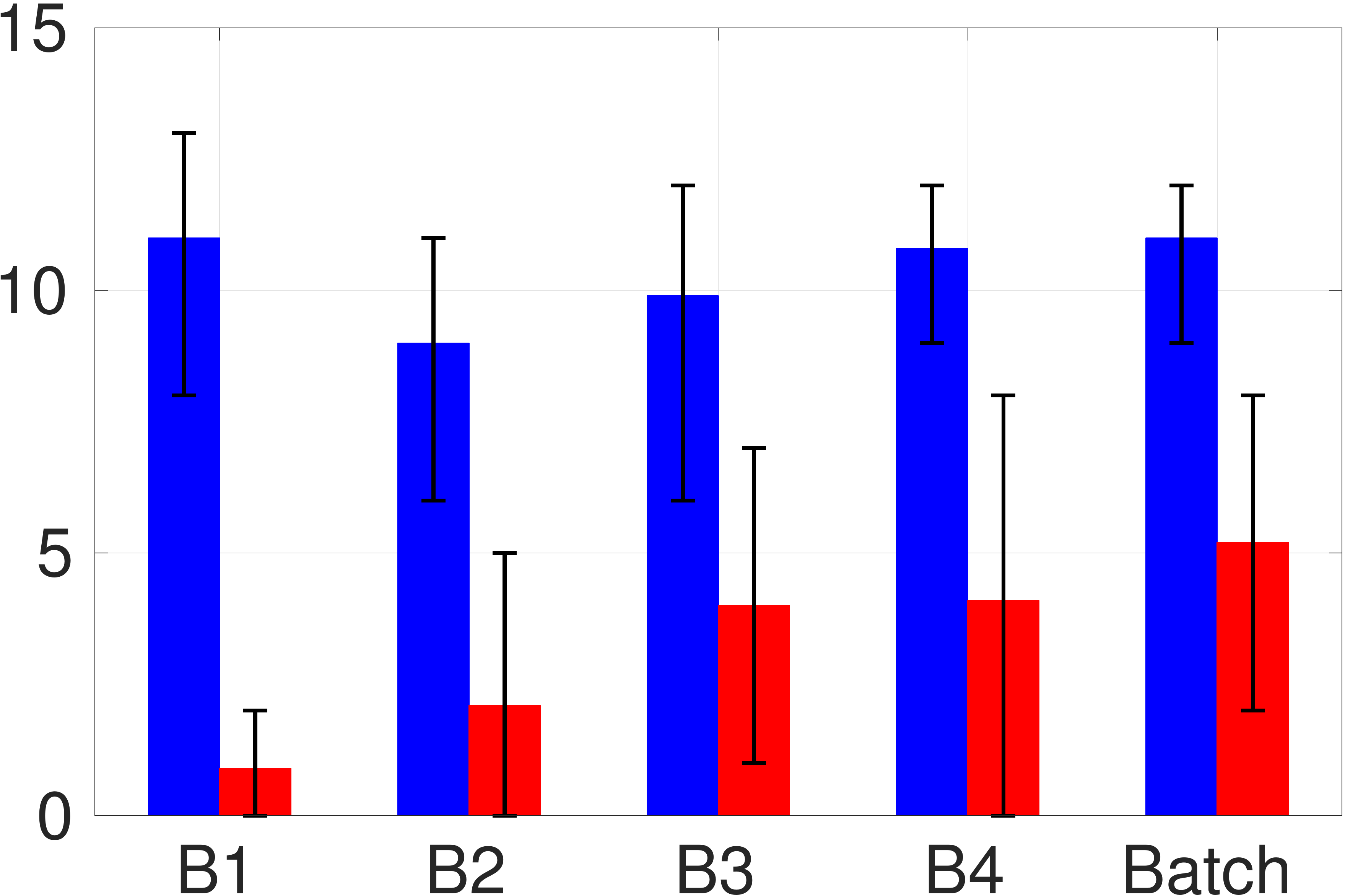}\label{fig:fig1_fourth_case}}
\caption{Deviations in detection of parents}
\label{fig:z}
\end{figure*}
\begin{figure*}[t]
\centering
\subfloat[{\em A}: $\mu_2=100$, $\kappa_2=40$]{\includegraphics[width=.21\textwidth]{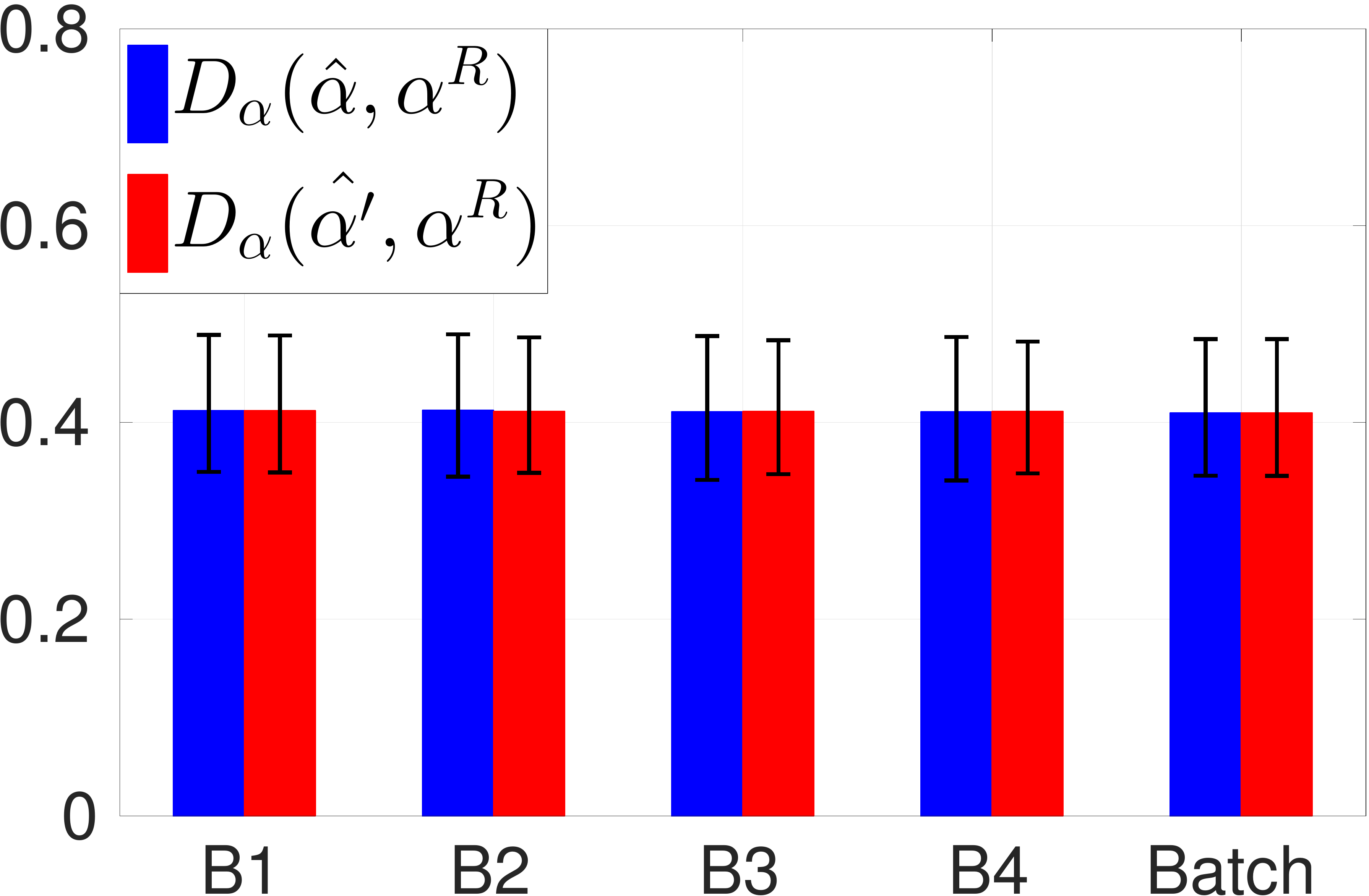}\label{fig:fig1_first_case}}\hspace*{0.17in}
\subfloat[{\em B}: $\mu_2=100$, $\kappa_2=2$]{\includegraphics[width=.21\textwidth]{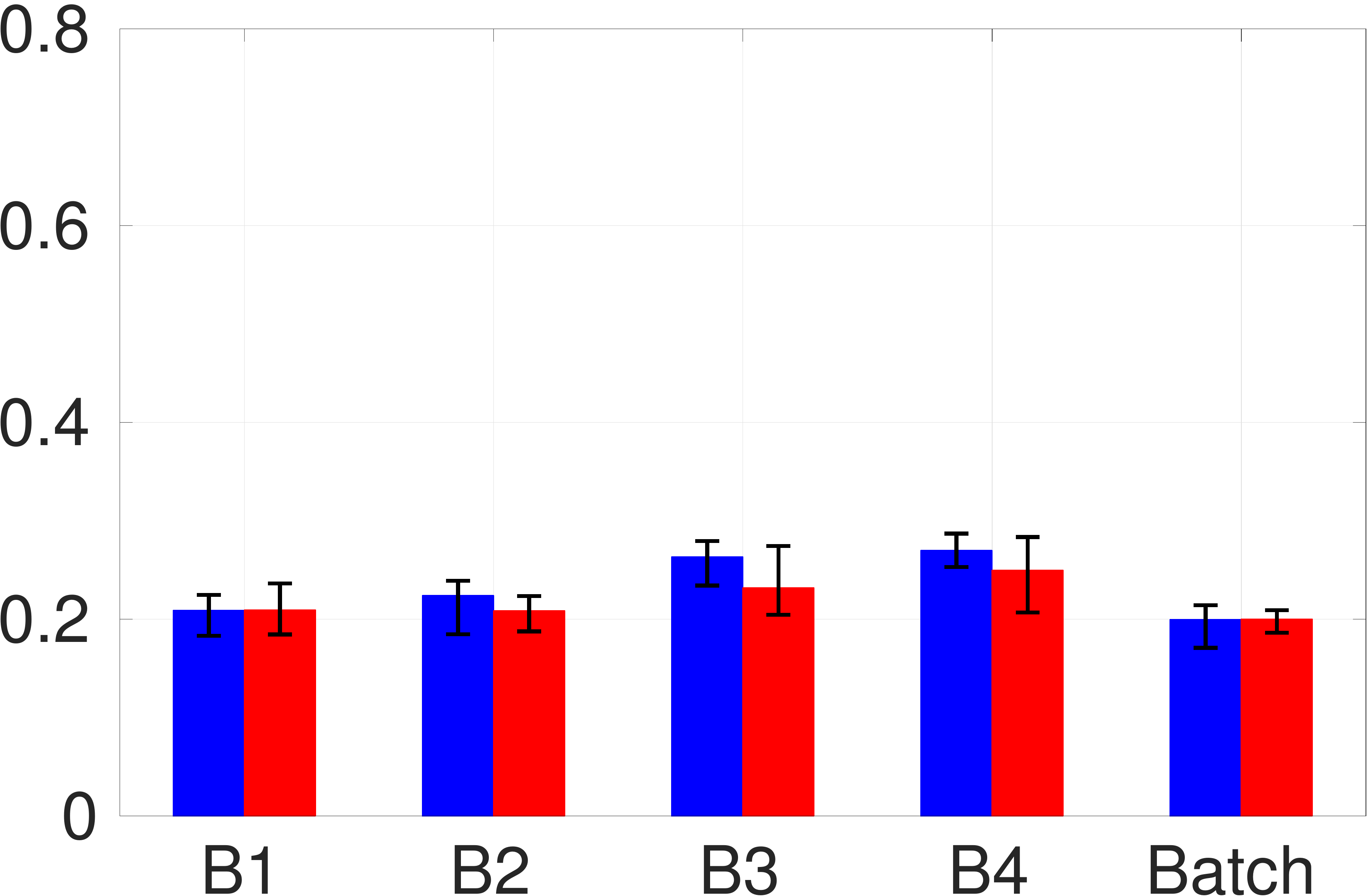}\label{fig:fig1_second_case}}\hspace*{0.17in} \subfloat[{\em C}: $\mu_2=11$, $\kappa_2=40$]{\includegraphics[width=.21\textwidth]{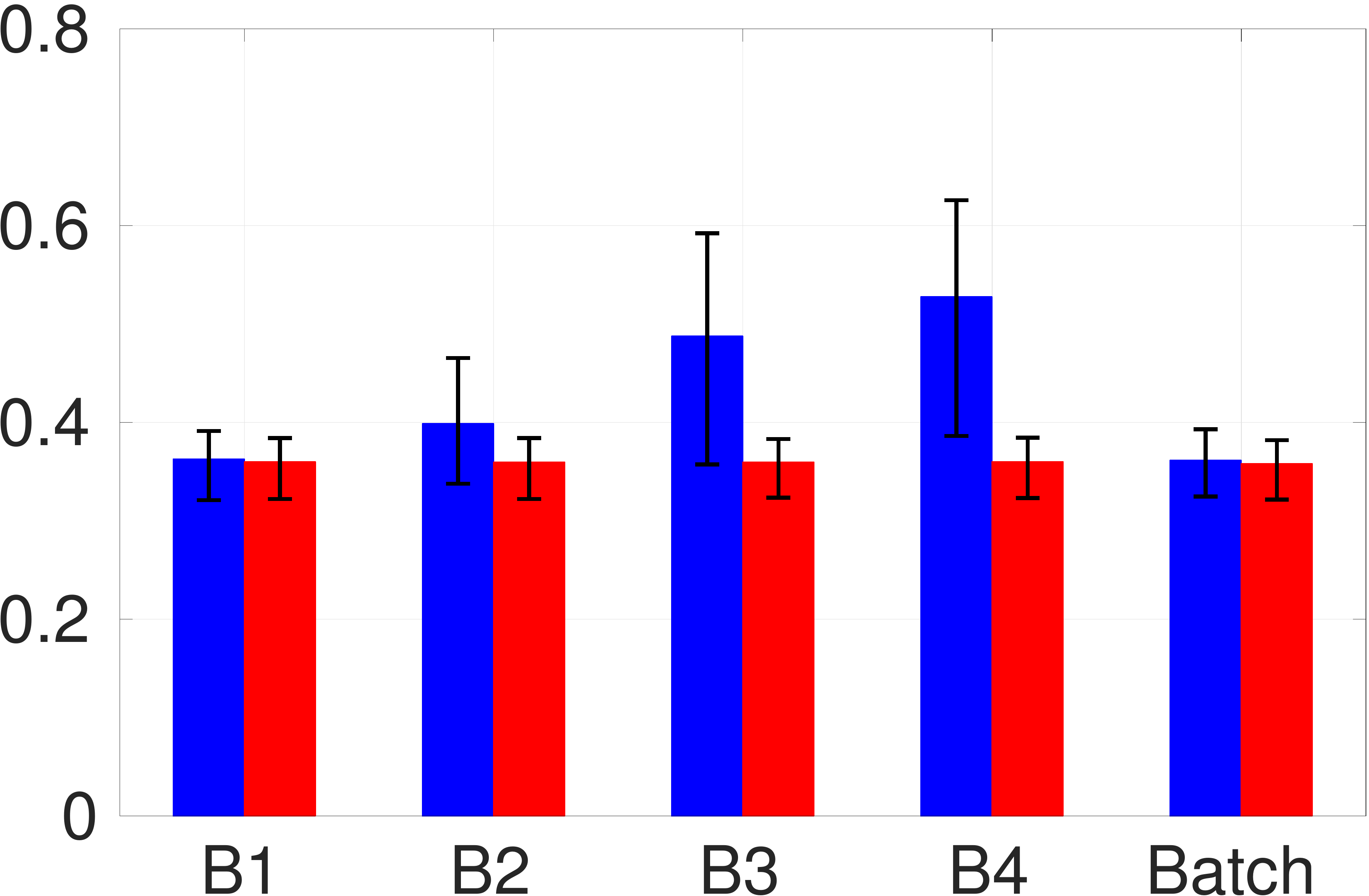}\label{fig:fig1_third_case}}\hspace*{0.17in}
\subfloat[{\em D}: $\mu_2=11$, $\kappa_2=2$]{\includegraphics[width=.21\textwidth]{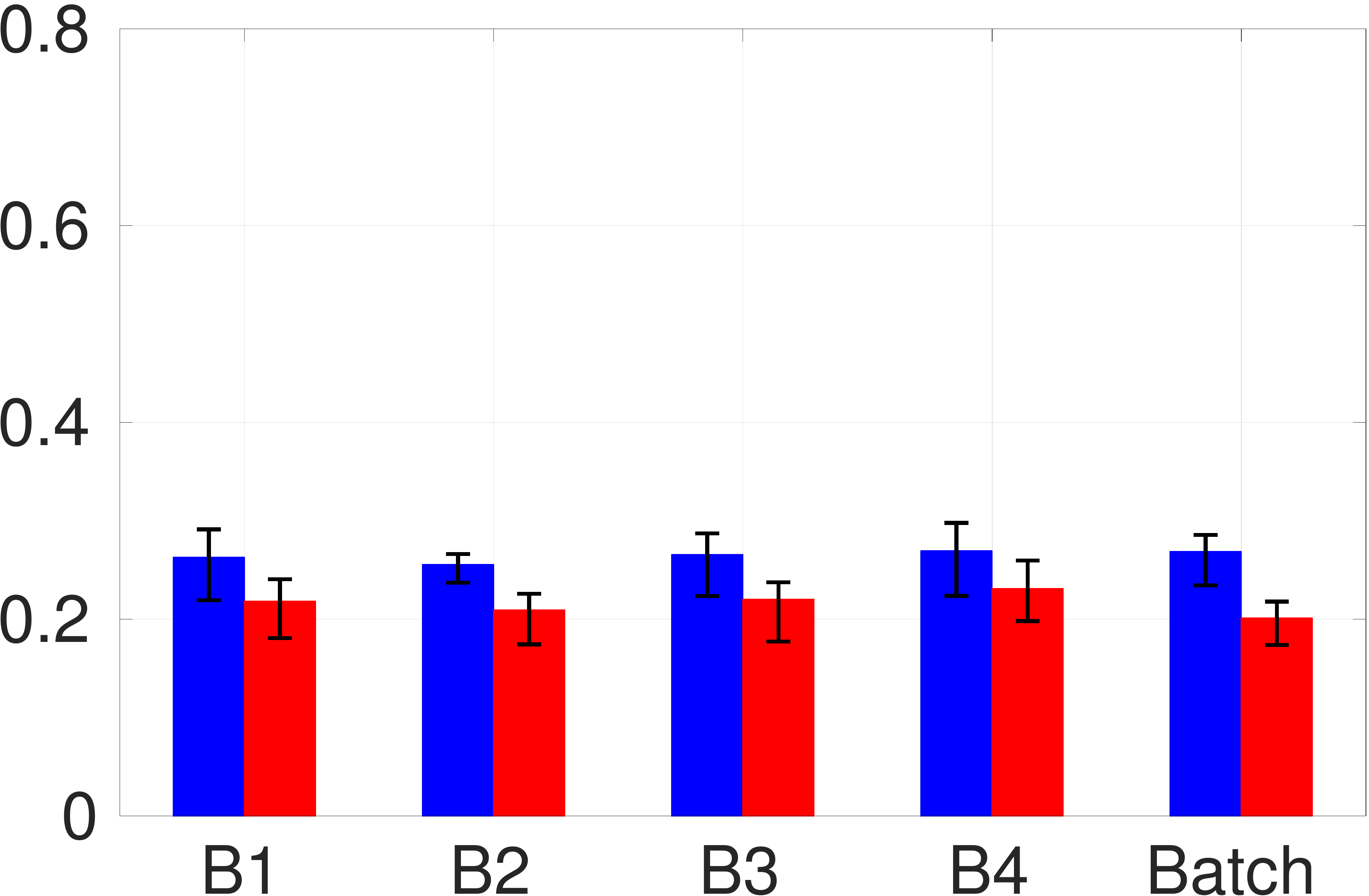}\label{fig:fig1_fourth_case}}
\caption{Deviations in detection of link strengths}
\label{fig:a}
\end{figure*}
\subsection{Synthetic Data}
Since we wish to evaluate the performance of the algorithms in
scenarios where the modelling assumptions are exactly met, we generate
a dataset based on the model described in Section \ref{model}. We
assume that node $1$ is the source of the infection.  For each node
$i>1$, we randomly construct $\pi^i $ by including each $j\in
\{1,\dots,i-1\}$ as a potential parent for $i$ (i.e., $j \in \pi^i$)
with probability $0.5$. Then, we uniformly choose $j \in \pi^i$ for
each node $i$ and build a random directed tree $\cT$ with edges $j
\rightarrow i$. We assign the link strength matrix
$\balpha^R=[\alpha^R_{ij}]_{N\times N}$ to this model and call them the {\it
  true link strengths}. $\alpha^R_{ij}\neq 0$ if and only if $j\in
\pi^i$. In this case, $\alpha^R_{ij}$ is drawn from a gamma distribution
$\Gamma (\kappa_1,\theta_1)$ if $ j \rightarrow i \in \cT$ and from $\Gamma
(\kappa_2,\theta_2)$ if $ j \rightarrow i \notin \cT$. We choose the
parent of node $i$ from all the nodes $j \in \pi^i$ based on a random
sampling with weights $\alpha^R_{ij}$. These parents are called
\textit{true parents} and are denoted by
$\bz^R=[z_1^R,\dots,z_N^R]^T$. Knowing the values of $z^i$ and
$\alpha^{iz^i}$, we then generate the \textit{true infection times}
$\bt^R=[t_1^R,\dots,t_N^R]^T$ based on the geometric distributions
described in \eqref{prior_t}. After choosing the random set of true
diffusion parameters $( \bz^R,\bt^R,\balpha^R )$, we
choose the length of time series $N_T$ to be 10 samples more than the
maximum infection time (i.e., $N_T=10+\max_{i} t_i^R$). Finally, we
generate data signals $\bd=\{ \bd^1,\dots,\bd^N\}$ of length
$N_T$ based on two different Gaussian distributions for before and
after being infected. Hyperparameters $\gamma_1^i=(\mu_1,\sigma_1)$
and $\gamma_2^i=(\mu_2,\sigma_2)$ are used for all nodes $i \in \cN$.

We aim to investigate whether incorporating the infection diffusion
model can improve the estimation of infection times compared to
univariate changepoint estimation. Hence, we compare two estimates of
infection times: the MAP estimate $\hat{\bt}$ based on the model and
inference techniques described in this paper; and an estimate
$\hat{\bt'}=[\hat{t'}^1,\dots,\hat{t'}^N]^T$ derived by maximizing the
likelihood of infection time for individual time-series, i.e,
$\hat{t'}^i= \arg \max_{t}
f(d_{1:t}^i;\gamma_1^i)f(d_{t+1:N_T}^i;\gamma_2^i)$.

We examine batch and online Bayesian inference algorithms  in a network of $N=20$ nodes. For the online setup, data signals are divided into four blocks of equal length ($\bB_1$ to $\bB_4$), i.e., $M =\lfloor \frac{N_T}{4} \rfloor$. We test inference algorithms in four scenarios. In all of the scenarios, $\mu_1=10$,
$\sigma_1=\sigma_2=1$, $\kappa_1=1$, $\theta_1=\theta_2=0.5$, and
$r_b=0.5$. In the first scenario (Scenario $A$), $\mu_2=100$ and
$\kappa_2=40$. Hence, not only can the infection times be detected
with high likelihood, but also the links that demonstrate the parental
relationships can be easily distinguished from the rest of the links
due to their high average strengths. In Scenario $B$, $\mu_2=100$ and
$\kappa_2=2$. Therefore, the difference between parent and non-parent
links are not significant and the underlying network is not easily
detected even though infection times are still recognizable with high
likelihood. The same trend exists in Scenarios $C$ and $D$ except for
the fact that in these two scenarios the means of two Gaussian
distributions are close i.e. in Scenario $C$, $\mu_2=11$,
$\kappa_2=40$ and in Scenario $D$, $\mu_2=11$, $\kappa_2=2$.

Figure \ref{fig:t} shows the mean and $95\%$ confidence intervals of
deviations of the two estimates $\hat{\bt}$ and $\hat{\bt'}$
from the true infection times $\bt^R$ for 100 realizations of true diffusion parameters.
The deviation between two infection time vectors (e.g., $D_t(\hat{\bt},\bt^R)$) is defined as the average
absolute difference between infection times across all nodes. When the true infection time is greater than the time index of the end of a batch $b$ (i.e., $t_i^R>\bB_b[Mb]$), we set the true infection time to null for that batch. If either the estimated infection time of a node or its true value is null for a batch $b$, we replace the null value with the end of the batch when calculating the deviation metric. For each of the algorithms, $N_{MCMC}=10^5$ samples are generated, the
first $N_{burn}=10^3$ samples are discarded, and every
$N_{thin}=10^{\text{th}}$ of the rest of the samples are kept. The $i^{\text{th}}$ component of $\hat{\bt}$ denotes the most observed infection time for node $i$ among the stored samples while the
$i^{\text{th}}$ component of $\hat{\bt'}$ shows the maximum likelihood
estimation of infection time of node $i$ while ignoring other nodes
and the underlying network. 
As expected, the infection times are perfectly detected in the first
two scenarios even though in Scenario $B$, the link strengths don't
have significant difference between parent and non-parent links. As
opposed to Scenarios $A$ and $B$, deviations of detected infection
times from their true values are non-zero in Scenario $C$ and they are
even larger in Scenario $D$ where neither infection times nor parental
relationships are easy to detect. However, we see that for both batch
and online inference methods, exploiting the underlying diffusion
network in detecting the infection times leads to more accurate
results in average.

Next, we wish to investigate how not knowing the infection times
affects the performance of detecting the parental relationships and
their strengths. In order to do this, we compare our setup in which all
the infection parameters are unknown with a setup in which infection
times are perfectly known and $\bz$ and $\balpha$ are the only
parameters to be detected. The MAP estimates for the case of unknown
infection times are denoted by $\hat{\bz}$ and $\hat{\balpha}$. We use
the same Gibbs sampling method to perform inference for the case when
the infection times are known (except there is no longer a need to
sample infection times). The MAP estimates for this latter case are
denoted by $\hat{\bz'}$ and $\hat{\balpha'}$. We calculate the error
metric for parent identification, 
$D_{z}(\hat{\bz},\bz^R)$, as the average number of nodes
whose parents are different in $\hat{\bz}$ and $\bz^R$.
Similarly, $D_{\alpha}(\hat{\balpha},\balpha^R)$ denotes the deviation
between detected and true link strength values and is equal to the
average over all links of the absolute difference between estimated and true $\alpha_{ij}$.

\begin{figure}[hb]
\centering
\subfloat[{\em C}: $\mu_2=11$, $\kappa_2=40$]{\includegraphics[width=.21\textwidth]{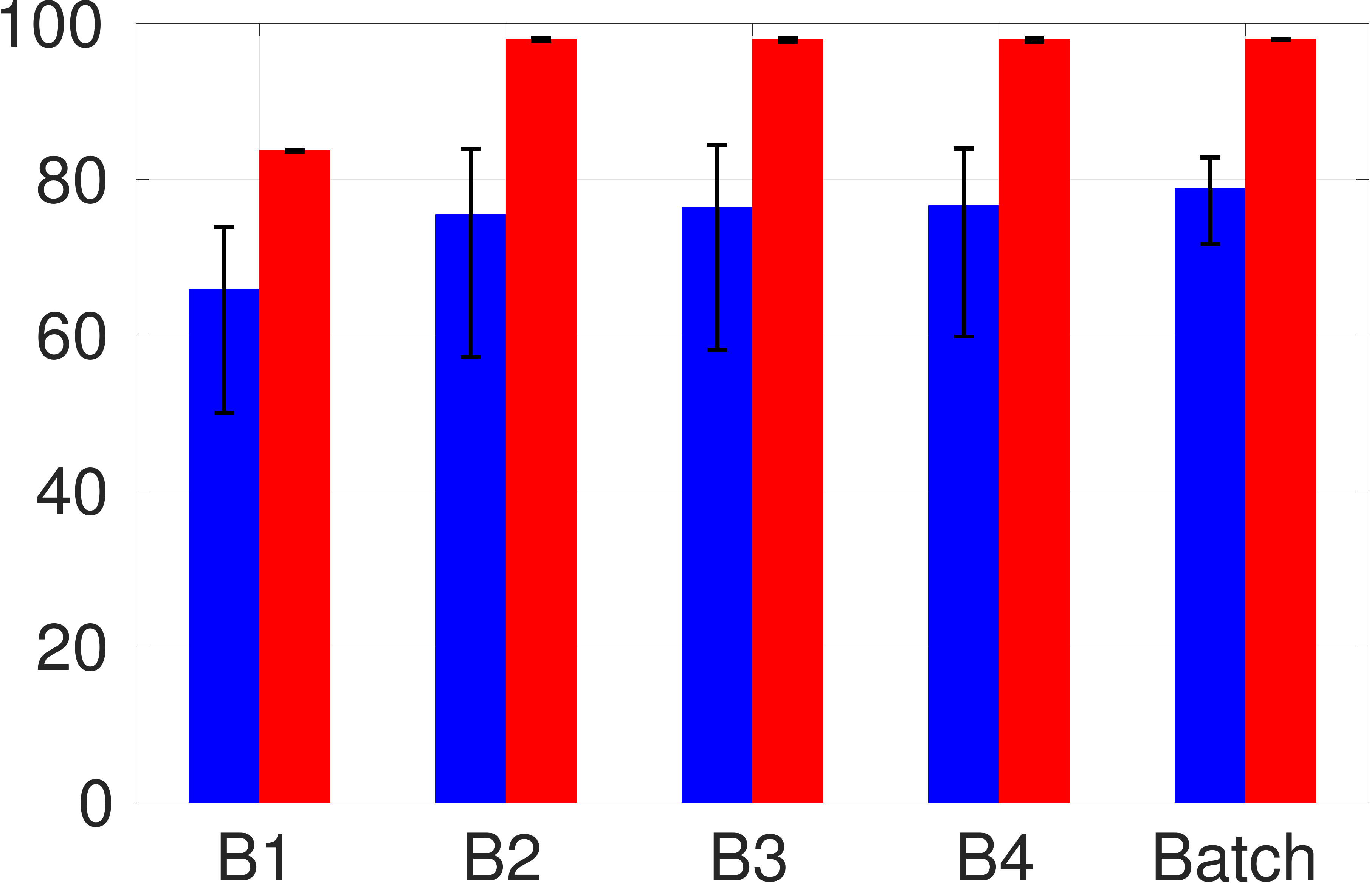}\label{fig:percent_z_1}}\hspace*{0.17in}
\subfloat[{\em D}: $\mu_2=11$, $\kappa_2=2$]{\includegraphics[width=.21\textwidth,trim={0 7cm 0 7cm},clip]{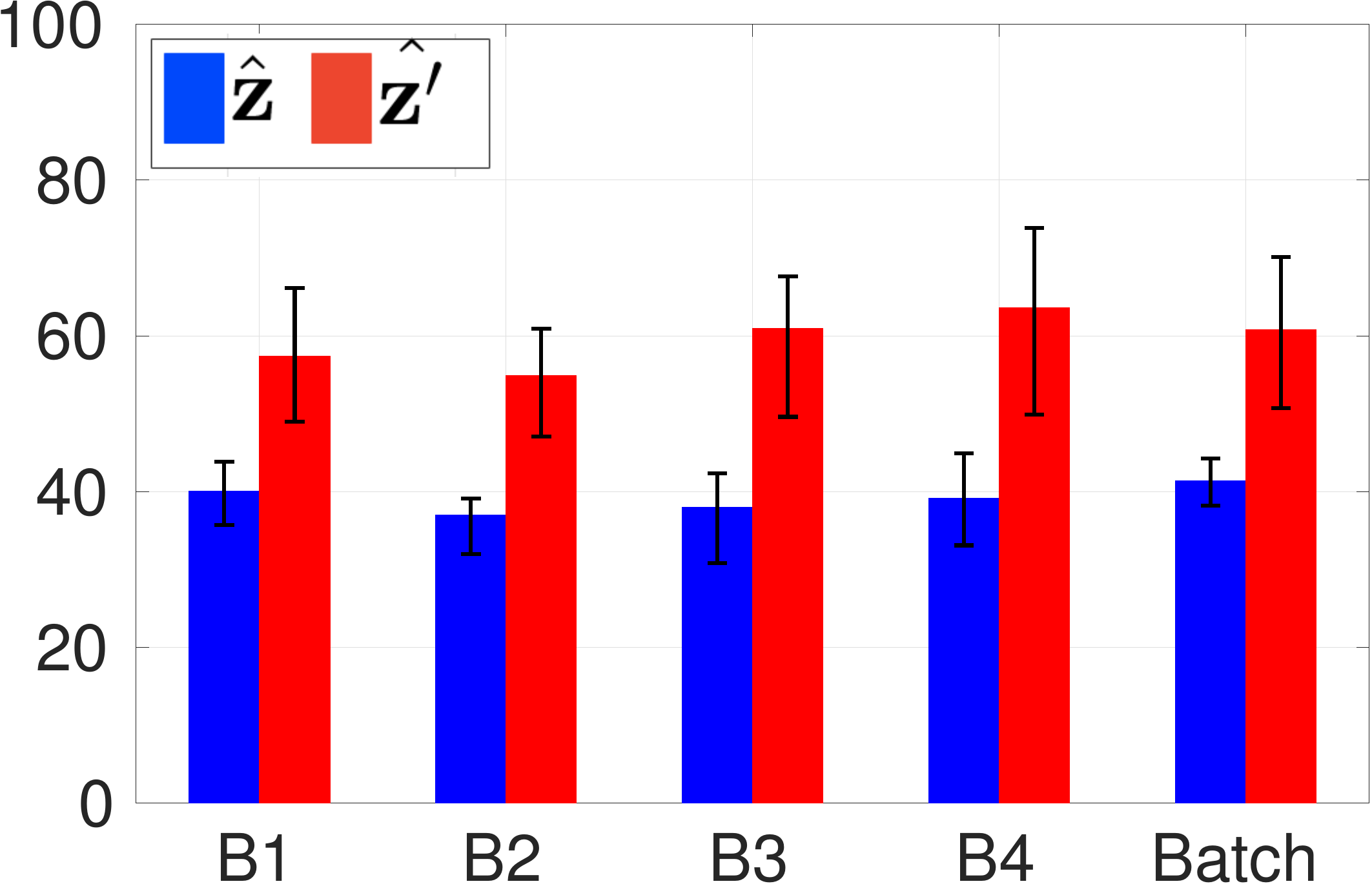}\label{fig:percent_z_2}}
\caption{Average percentage of samples identifying the correct parents in Scenarios C and D }
\label{fig:percent_z}
\end{figure}

Figures \ref{fig:z} and \ref{fig:a} show the error metrics for the
estimation of parents and link strengths for the batch and online algorithms.
In Scenarios $A$ and $B$, when infection
times are easy to infer, the knowledge of infection times provides
little benefits and the error metrics are approximately the same.
However, when the infection times are hard to detect (Scenarios $C$
and $D$), the deterioration in the estimation accuracy is more observable. Figure \ref{fig:percent_z} shows the average percentage of samples identifying the correct parents in Scenarios C and D.
\subsection{Measles and Chickenpox Data}

We study the number of weekly reported cases of measles and chickenpox
in England and Wales as shown in Figure \ref{fig:map1}. The analysis
is based on the data from seven large cities (London, Bristol,
Liverpool, Manchester, Newcastle, Birmingham, and Sheffield), with
populations ranging from $300,000$ to $10$ million (out of a total
population of approximately $50$ million). The primary infections
occur every two years in the period from September to December of the
following year. We focus our analysis on these time windows.

Our model for this data assumes that an infection commences in one
city and then by movement of individuals between two cities emerges in one
of the other cities. The model does allow for the possibility of
infections arising spontaneously due to unrelated influences in
multiple cities. In such a case, multiple nodes (cities) have no
parent in the inferred model.  This can capture scenarios in which
infections are caused by interactions with other cities not included
in the model (e.g., visitors from other countries). It can also
explain the case when the residual infection in a city gives rise to a
new outbreak, with no disease transfer from another city.

Quantitative studies
(e.g.~\cite{bolker1996impact}) have explained the decrease in the
number of reported cases in the data after widespread use of a vaccine began in
$1968$. Also, studies such as \cite{bolker1993chaos} have justified
the biennial data peaks, claiming that they were caused by exhaustion
and subsequent build-up of susceptibles in the population as well as
seasonal changes in virus transmission. Finally,
\cite{wang2013modelling} shows that once an infection starts in a
region, the number of hospitalized cases can be approximated by a
log-normal function of the number of days that have passed since the
infection began. The number of
hospitalized cases in region $i$ at the $n^{\text{th}}$ week is
denoted by $h_n^i$ and defined as the cumulative number of cases minus
the cumulative number of deaths and recoveries. Hence, for $n\geq
t^i_1$ we have
\begin{equation}
h_n^i=\frac{A_h^i}{\sqrt{2\pi}\sigma_h^i 7(n-t_1^i)} \exp(-\frac{(\ln[7(n-t_1^i)] -\mu_h^i)^2}{2{\sigma_h^i}^2}).
\end{equation}
Here $A_h^i$ is a constant, $\sigma_h$ is the variance parameter of
the log-normal, and we set $\mu_h^i=\ln D_h^i+{\sigma_h^i}^2$, where
$D_h^i= \arg \max_n h_n^i$. The coefficient $7$ is due to the fact
that the data is reported weekly.

For each time-series, we use the log-normal function to model the data
after an infection. For a candidate infection time, we find
$\sigma_h^i$ and $D_h^i$ such that the Mean Square Error (MSE) is
minimized. We also choose $A_h^i$ such that
$d^i_{D_h^i}=h^i_{D_h^i}$. The ``individual infection time estimate''
is then the candidate infection time that has minimum MSE after the
log-normal fit. We model the residual using a
Gaussian distribution. We also assume that the data follows a normal distribution
before the individual infection times, i.e.,
\begin{equation}\label{chickenpox_normal}
d_n^i|t_1^i \sim \begin{cases}\cN(\mu_1^i,\sigma_1^i) \quad & \text{if } n> t_1^i,\\
\cN(h_n^i+\mu_2^i,\sigma_2^i)\quad & \text{if } n \leq t_1^i.
\end{cases} 
\end{equation}
In these models, we use empirical variances derived from the previous time-period.

\begin{figure}[htb]
\centering
\subfloat[Reported cases for Liverpool in Sep 1958 to Dec 1959]{
\includegraphics[width=0.46\textwidth]{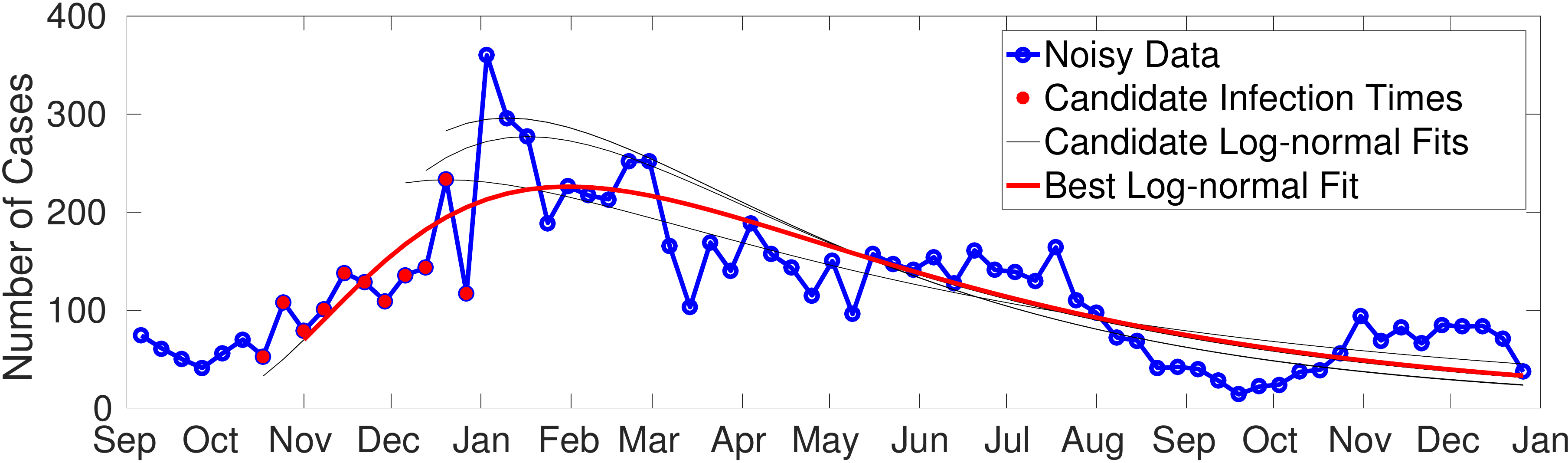}\label{fig:qqs1}}\\
\subfloat[Reported data before the individual infection time]{
\includegraphics[width=0.215\textwidth]{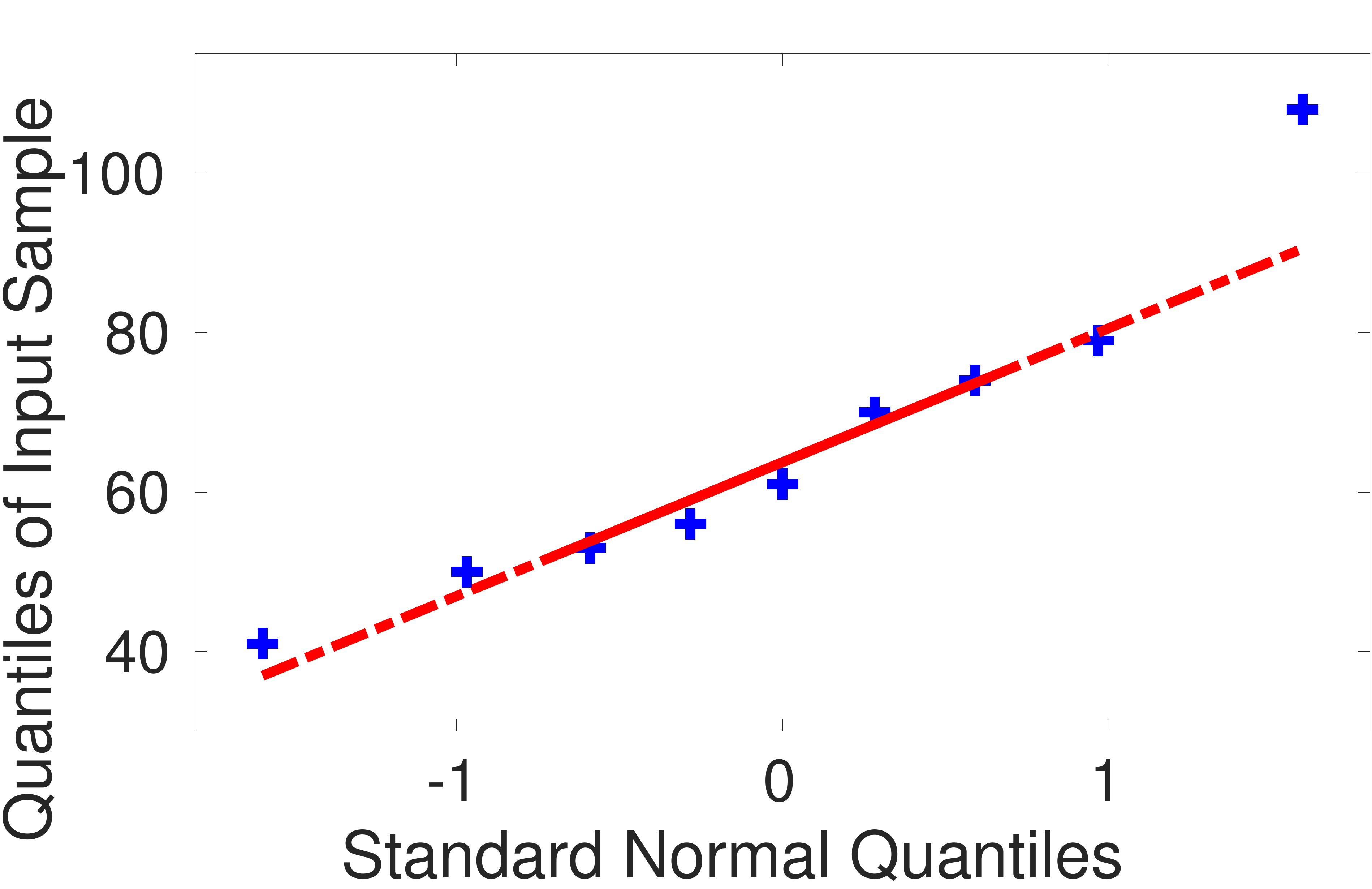}\label{fig:qqs2}}\hspace{1mm}
\subfloat[Residual noise after the individual infection time]{
\includegraphics[width=0.215\textwidth]{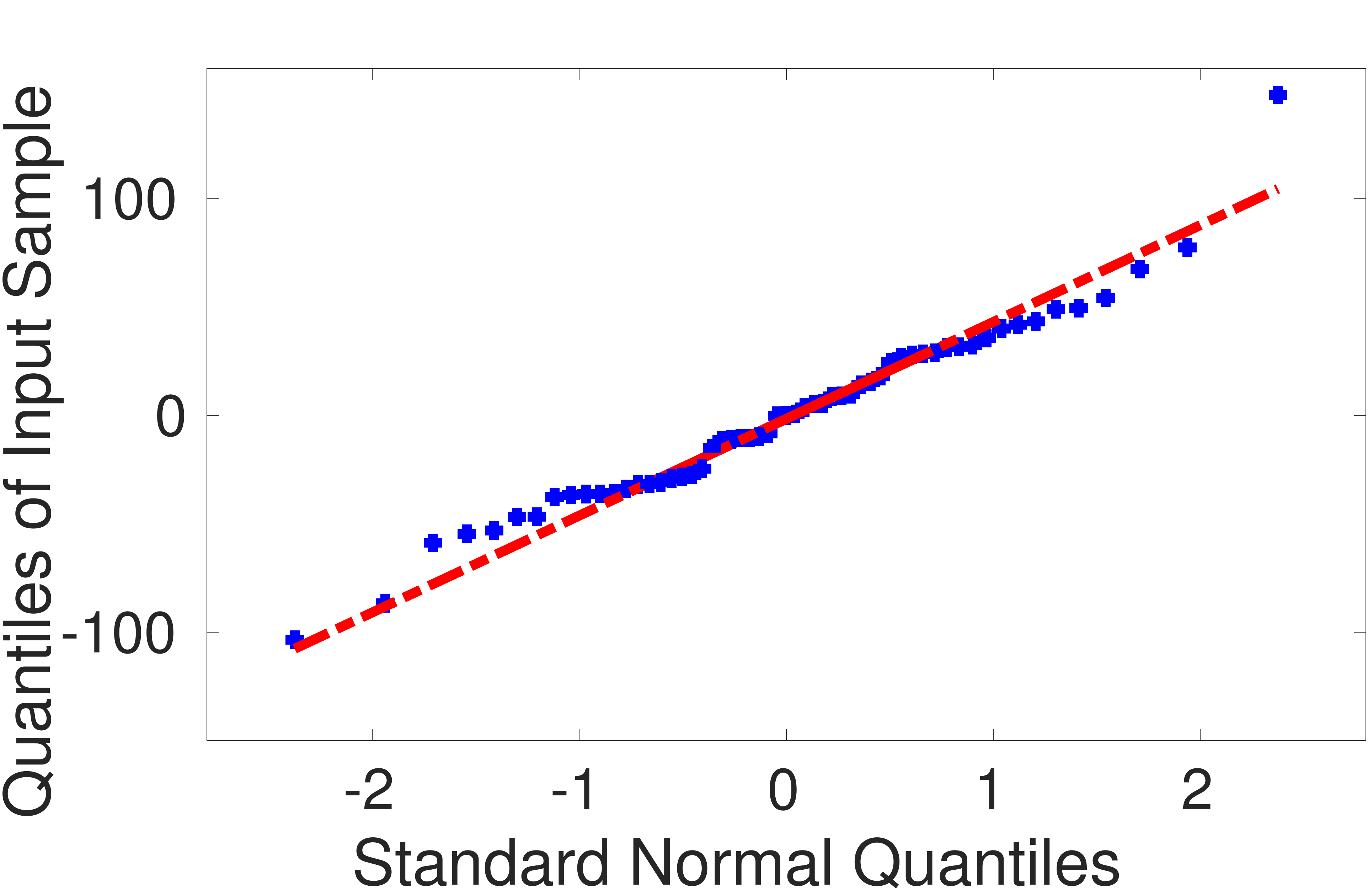}\label{fig:qqs3}}
\caption{(a) Reported cases time series with best fit log-normals for
  different candidate infection times, (b) Quantile-quantile (qq) plot
  of the reported data before the individual infection time versus
  standard normal distribution, (c) qq plot of the residual noise
  after the individual infection time versus standard normal
  distribution}
\label{fig:qqs}
\end{figure}

Figure \ref{fig:qqs} shows an example for 1958-1959 in Liverpool. The
candidate infection times are shown by red points in Figure
\ref{fig:qqs1}. In \ref{fig:qqs1}, the log-normal fits for
each candidate infection time are shown in black. For the
example of \ref{fig:qqs1}, the individual infection time estimate is November
$1^{\text{st}}$. The quantile-quantile plots in Figures~\ref{fig:qqs2}
and~\ref{fig:qqs3} provide support for the Gaussian model. Results for
other years are provided in~\cite{techrep2016}.

\subsubsection{Estimating Infection Times}

Figure \ref{fig:series} shows the reported data for the 1958-1959 study period. The individual infection time estimates are indicated
by the change of color from green to red. In all the four successive biennial study periods from 1956 to 1963,
Liverpool (node $3$) has the earliest individual infection time
estimate. We consider this node as the source of the infection. Given
the entire data for each study period, we intend to infer the
underlying network as well as the infection times. We use the prior
distributions proposed in \eqref{prior_t}, \eqref{prior_z}, and
\eqref{prior_alpha}. The data is modelled as
in~\eqref{chickenpox_normal}, with $h_n^i$ calculated for each
candidate infection time by minimizing the MSE as explained above. In
each study period, we choose the hyper-parameters
$\mu_1^i,\sigma_1^i,\mu_2^i,\sigma_2^i,\kappa,\theta$ based on the
reported data and the individual infection time estimates for the
preceding period.

Any node $j$ whose individual infection time estimate is earlier than
the individual infection time estimate of node $i$ in a study period
is considered to be a potential parent, i.e. $j \in \pi^i$, for the
next study period. We generate $N_{MCMC}=10^5$ samples, discard the
first $N_{burn}=10^3$, and store one out of every $N_{thin}=10$ of the
rest. The means of the retained samples are used to calculate estimated
infection times. These estimated infection times are shown by the black
vertical lines in Figure \ref{fig:series}. The estimated
infection times are quite close to the individual estimates.

\subsubsection{Predicting Infection Times}
\begin{figure}[htb]
  \includegraphics[width=0.49\textwidth, height=0.46\textwidth]{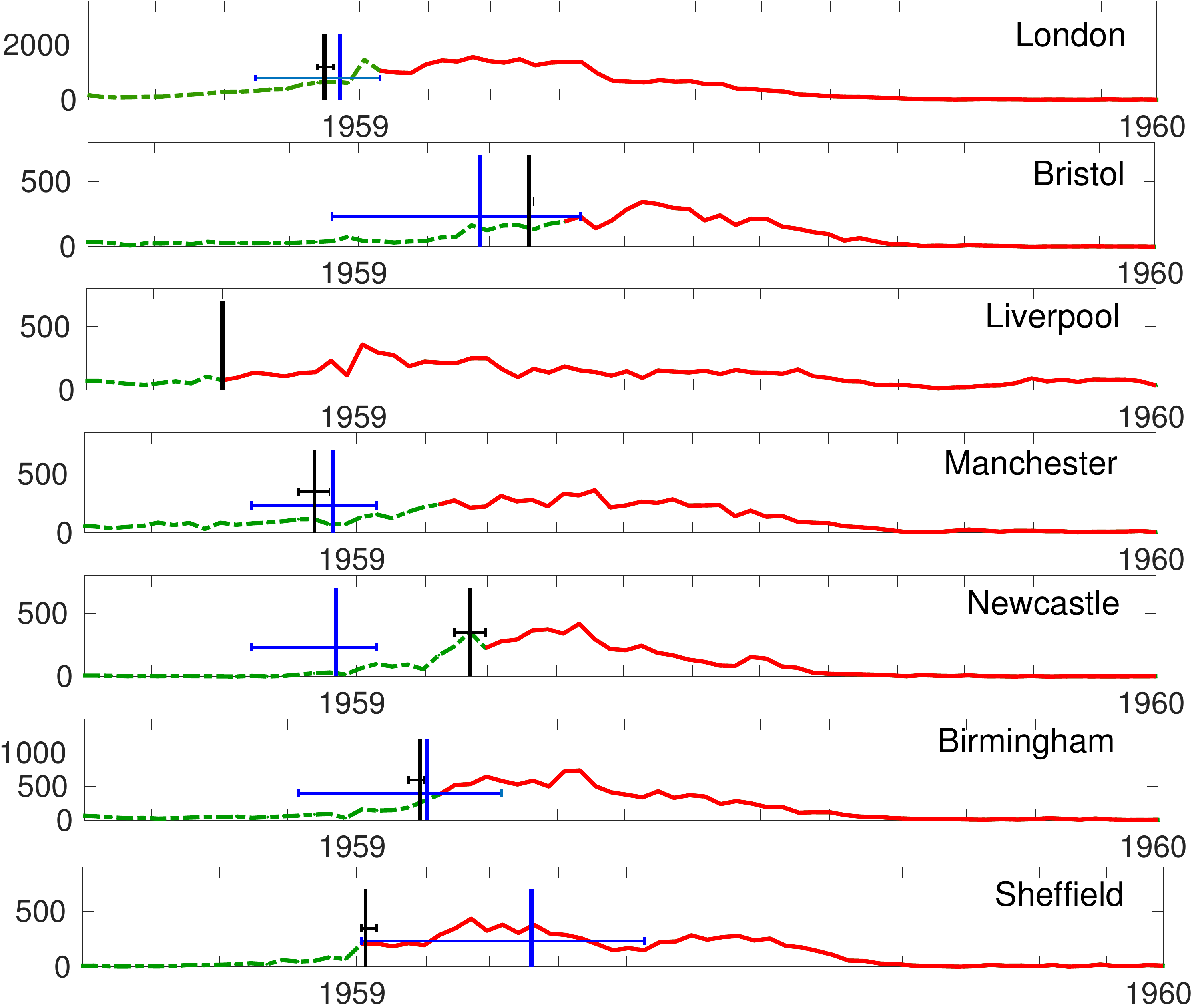}
  \caption{Number of reported cases with estimated and predicted infection times in 1958- 1959. dashed green line: susceptible state, solid red line: infected state (after individual infection time), black lines: mean and 25-75\% of estimated values, blue lines: mean and 25-75\% of predicted values}
  \label{fig:series} \end{figure}

\begin{figure}[ht]
\centering
\includegraphics[width=0.35\textwidth]{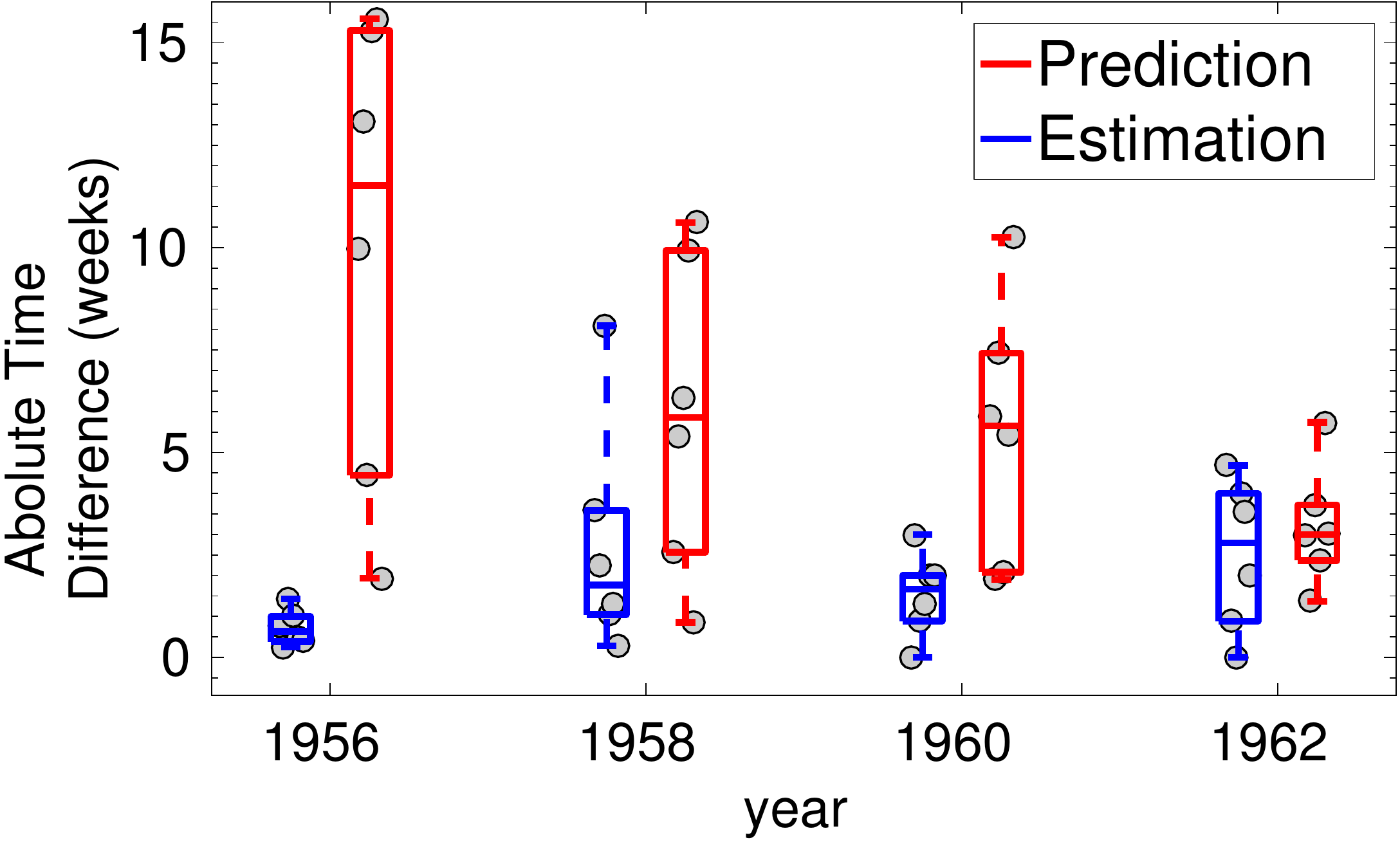}
\caption{The absolute difference between estimated (predicted) infection times and the individual ones. The central mark of each box is the median, the edge of the box are the 25th and 75th percentiles, the whiskers extend to the highest values not considered as outliers.}
\label{fig:boxplot}
\end{figure}

In addition to estimating the infection times, we are interested in
predicting them beforehand without having access to the reported
data. After we have learned a model, we can form a prediction using
only the (individual) infection time estimate of the source node. This
allows us to detect the onset of the infection in Liverpool and use it
to predict when infections will arise in London and Manchester, for
example.  We use the inferred network structure for each study period
to predict the infection times of the next period. For example, the
infection times for 1958-1960 are predicted using the model learned by
processing 1956-1958 data. For every stored sample from the
distribution, we predict the infection time of node $i$ by
$t_1^{z_1^i}+\delta^{iz_1^i}$ where $\delta^{iz_1^i}$ is the mean of a
geometric distribution with parameter $1-e^{-\alpha_1^{iz_1^i}}$. The
mean and $25-75 \%$ confidence intervals of these predicted values are
respectively shown by the vertical and horizontal blue lines in Figure
\ref{fig:series}. Figure \ref{fig:boxplot} shows the absolute
difference between the average estimated (predicted) infection times
with the individual infection time estimates for all nodes except for
the source. 

\subsection{Earthquake Data}
\begin{figure}[ht]
\centering
\includegraphics[width=0.48\textwidth, height=0.32\textwidth]{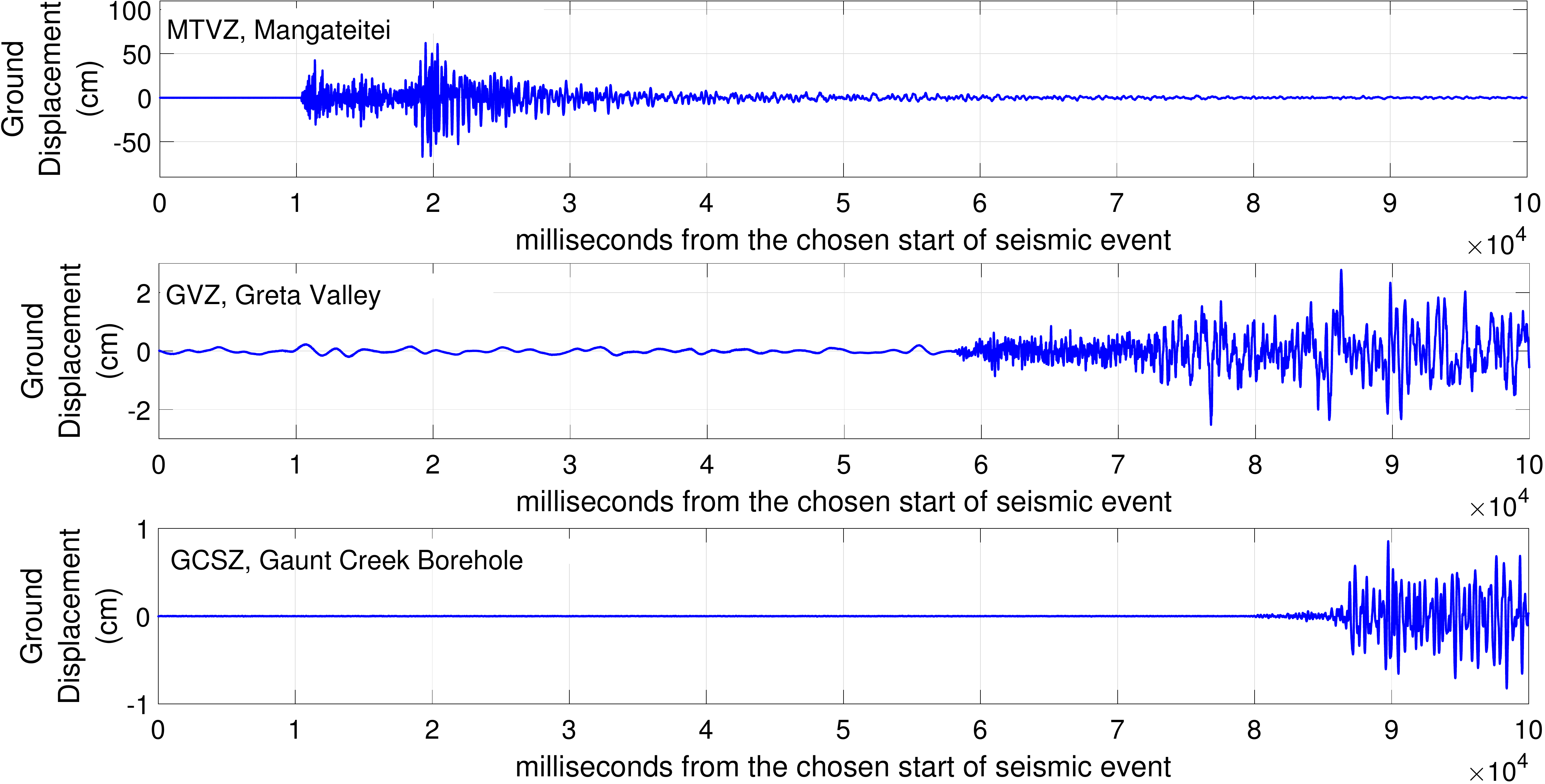}
\caption{Three recorded seismic waveforms for an earthquake happened on November 1st 2015 (Event ID: 2015p822263)}\label{fig:q_series}
\end{figure}

In addition to estimating the infection times, our proposed inference
approach can be used to detect the underlying network structure. As an
example, we use our proposed diffusion framework to study seismic
events in different regions of New Zealand\footnote{Publicly available
  at http://www.geonet.org.nz/quakes}. Seismic waves are energy waves
generated by earthquakes, volcanic eruptions, and other sources of
earth vibration. They travel along layers of the earth and across the
surface. Each wave is generated in one location and propagates out to
other regions. The {\em epicenter} of a seismic event is the location on the
surface of the earth directly above the cause of the wave. Our goal is
to locate the epicenter of a seismic event and compare it with the
reported real location. A seismograph is a device that records
earthquake waves and we consider the measuring stations equipped with
seismographs as the nodes of the diffusion network. We approximate the
seismic event as a propagation of energy waves between these discrete
nodes. A seismogram, the graph drawn by a seismograph, is a record of
the ground motion as a function of time. The recorded seismograms are
used as the observed data signals in our diffusion model. Three
examples are shown in Figure \ref{fig:q_series} for an earthquake that
occurred on November 1st 2015.

We choose the prior distribution for link strength values by fitting a
Gamma distribution to inverse values of the geographic distance
between station pairs. We also assume that seismic waves follow two
different Gaussian distributions before and after being infected. This
ignores the oscillatory structure and correlations in the time series,
but is sufficient for the estimation of the arrival of the seismic
wave. We denote the individual changepoint estimate of waveform $i$ by
$\hat{t'}^i$. Denoting the velocity of the seismic waves in the
related depth of the earth by $v$, we define the set of potential
parents for node $i$ as $\pi_i=\{j \in
\cN|\hat{t'}^j+\frac{D_{ij}}{v}<\hat{t'}^i\}$ where $D_{ij}$ is the
distance between stations $i$ and $j$. More precisely, node $i$ can
only be infected by node $j$ if the time difference between their
individual changepoints is larger than the time required for the
seismic wave to traverse their spatial distance. The precise speed with which seismic waves travel throughout the earth depends on several factors such as composition of the rock, temperature, and pressure. They typically travel at speeds between $1$ and $14$ $\frac{Km}{s}$. In our analysis, we
set $v$ to $13 \frac{Km}{s}$.

Since we do not know where the source of the infection is, we assume
that there is a dummy node within a radius of $D_{i0}$ of each real
node. These dummy nodes are candidate sources of the infection and
each is a potential parent of its corresponding node. All dummy nodes become
infected at the same time, $\underset{i}{\min }
(\hat{t'}^i-\frac{D_{i0}}{v})$. As for the prior distribution of the
infection time of node $i$, we know that it is zero for values less
than $\frac{D_{iz_1^i}}{v}$. This probability increases for greater
times up to a certain point and then it monotonically increases. We
approximate this behaviour with a geometric distribution as
in~\eqref{prior_t}.

\begin{figure*}[ht]
\centering
\subfloat[{\em Event ID}: 2016p105478]{
\includegraphics[width=0.62\textwidth, height=0.4\textwidth]{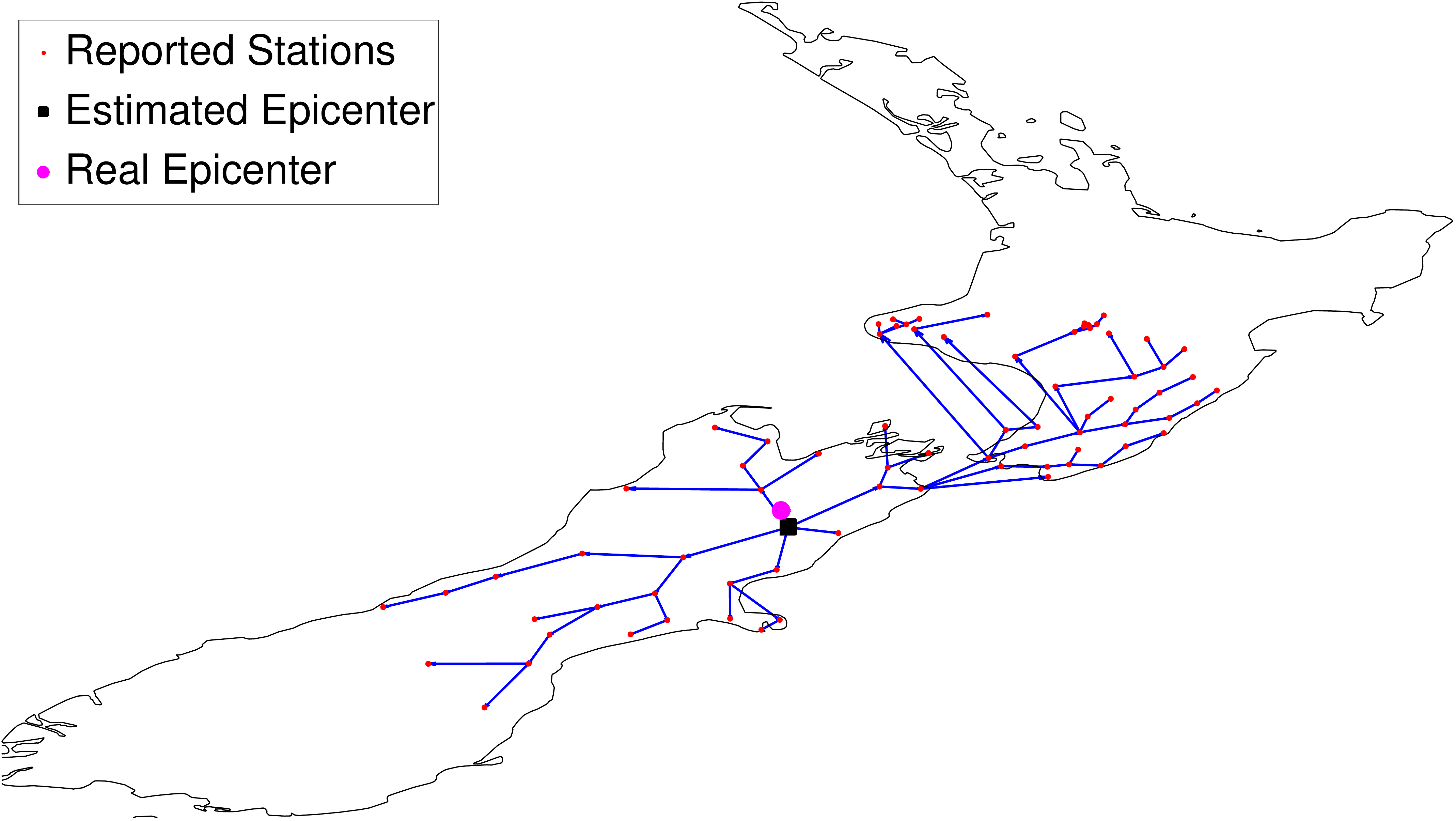}\label{fig:stations_1}}\hspace{-5cm}
\subfloat[{\em Event ID}: 2015p822263]{
\includegraphics[width=0.62\textwidth, height=0.4\textwidth]{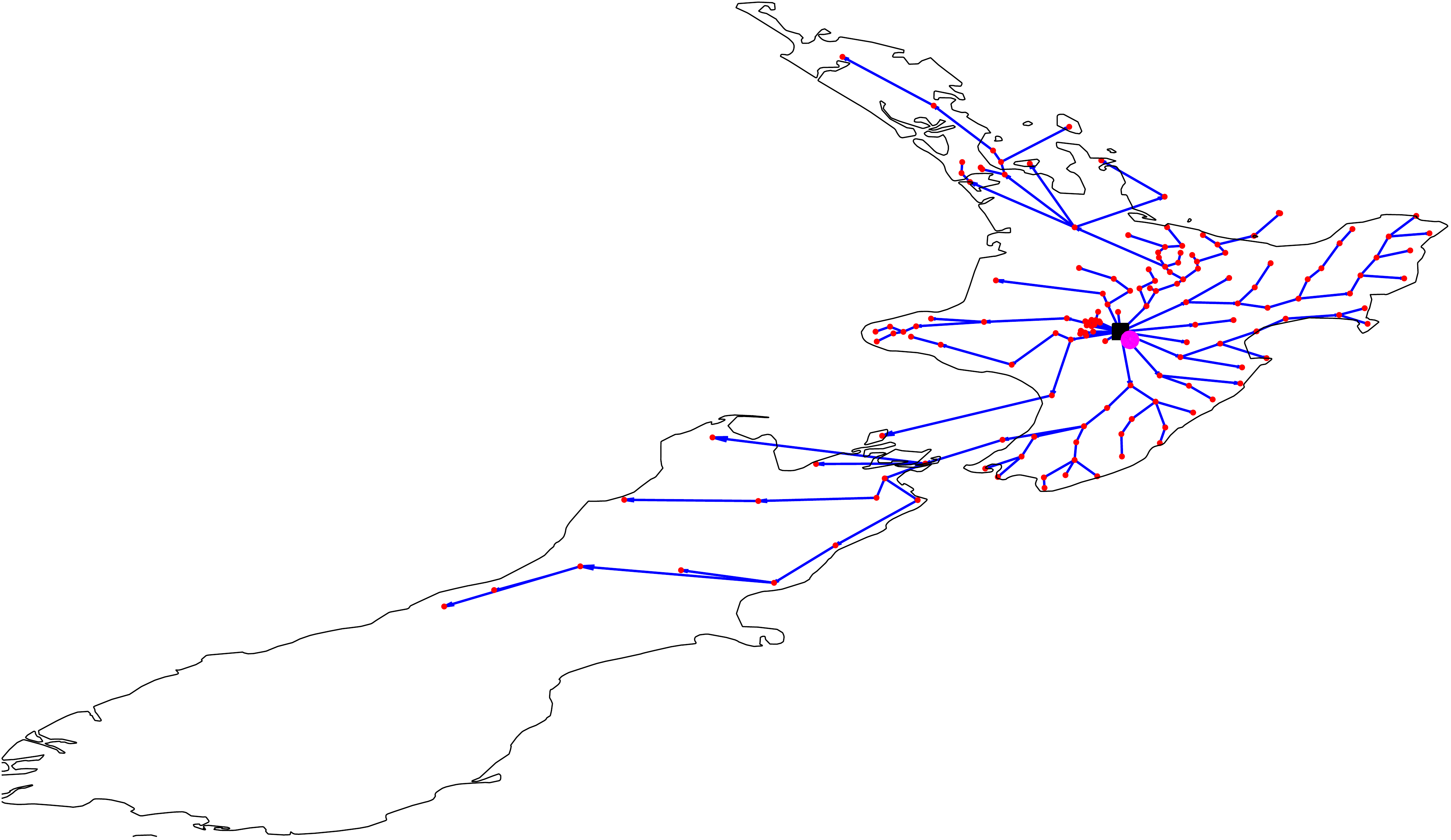}\label{fig:stations_2}}
\caption{Detected network structure for two seismic events in New Zealand, The distance between real and detected epicenters are (a) $29Km$ and (b) $15Km$ }
\label{fig:stations}
\end{figure*}

We run the batch inference algorithm on the diffusion network
with $N_{MCMC}=10^5$, $N_{burn}=10^3$, and $N_{thin}=10$. For each
node, we choose the node that marginally maximizes the posterior
distribution as its detected parent. This marginal MAP approximation
is the parent that occurs most often in the stored
samples. We consider the geographical midpoint of the nodes that are
infected by their dummy nodes as the approximate location of the
epicenter. The detected and real epicenters of two seismic events are
shown in Figure \ref{fig:stations} where $v=13 \frac{Km}{s}$ and
$D_{i0}=10 Km$ is used. We see that a tree-like network structure
exists where the root is close to the real epicenter of the seismic
event.

\section{Conclusion}\label{Conclu}
This paper presented a Bayesian framework for modeling the diffusion
of some sort of a contagion over a graph structure. We formulated the
diffusion process by introducing three main sets of parameters:
parental relationships, the strength of connections between node
pairs, and infection times. The main issue we addressed in this work
was to simultaneously infer these three sets of parameters using data
signals observed at each of these individual nodes. In order to do so,
we applied MCMC techniques to generate samples from the posterior
distribution of these parameters. One of the main contributions of
this paper is to address applications in which parameters must be
estimated before all of the data has been acquired. We addressed this
concern by developing an online version of the inference algorithm.

We evaluated the performance of our proposed inference approaches on
both synthetic and real world network scenarios. In the synthetic
dataset, model assumptions are exactly met and the ground truth is
perfectly known. Simulation results showed that considering the
underlying network structure in estimating infection times improves
accuracy compared to processing the data at each node individually. We
compared the precision of detecting the network structure (parental
relationships and link strengths) in scenarios with known and unknown
infection times. Finally, we tested our proposed inference algorithm
in practical scenarios. First, we showed how the algorithm can use the
number of reported cases of a contagious disease as the observed time
series to construct estimates and predictions of the time of the year
when disease outbreaks occur in a particular region. We then used the
inference approach to locate the epicenters of earthquakes using
the seismic waveforms recorded at seismic stations.
 
In future work, we aim to extend our methodology, developing more
sophisticated algorithms that account for more complicated infection
models, e.g., the {\it SEIR} (Susceptible-Exposed-Infected-Recovered)
model for observed time series. We also intend to improve our online
inference method so that it can process scenarios with dynamic
networks where nodes enter or leave the network or the set of
potential parents changes over time.

\appendices
\section{Proof of Proposition \ref{prop1}}\label{pp1}
\begin{proof}[\unskip\nopunct]
Replacing \eqref{h}-\eqref{h_z} in \eqref{integral}, we can
check that the integrals are separable for different nodes. Hence,
\eqref{integral} is equal to
\begin{equation}\label{integral1}
\begin{aligned}
&\prod_{i \in \cN}  \sum_{t_b^i \in \cS_1}h_t(t_b^i|\bx_{b-1};{t_{b}^{ML}}^i) \times \prod_{k \in \pi^i} h_{\alpha z}^k(t_b^i)\,,
\end{aligned}
\end{equation}
where 
\begin{equation}\label{h_az}
h_{\alpha z}^k(t_b^i)=\int_0^\infty h_{\alpha}(\alpha_b^{ki}|t_b^i,\bx_{b-1})\sum_{z_b^i \in \cS_2} h_z(z_b^i|t_b^i,\balpha_b,\bx_{b-1}) d_{\alpha_b^{ki}}\,.
\end{equation}
For each node $i$, we calculate the component in \eqref{integral1} for
two cases of $t_{b-1}^i=\phi$ and $t_{b-1}^i\neq \phi$
separately. When $t_{b-1}^i\neq\phi$, the integral is equal to
\begin{equation}
\begin{aligned}
\sum_{t_b^i\in \cS_1} &\delta(t_b^i-t_{b-1}^i) \times \\ &\prod_{k \in \pi^i}\int_0^\infty\sum_{z_b^i\in \cS_2} \delta(\alpha_b^{ik}-\alpha_{b-1}^{ik})\delta(z_b^i-z_{b-1}^i)d_{\alpha_b^{ik}}=1\,.
\end{aligned}
\end{equation}
When $t_{b-1}^i=\phi$, we start from the innermost integral of \eqref{h_az}
\begin{align}
\sum_{z_b^i \in \cS_2} &h_z(z_b^i|t_b^i,\balpha_b,t_{b-1}^i) \nonumber \\
&=\begin{cases}
1 &t_b^i=\phi\,,\\
1-\frac{\sum_{l \in \pi_b^i}\alpha_b^{il}}{\sum_{j \in \pi^i}\alpha_b^{ij}}+\frac{\sum_{l \in \pi_b^i}\alpha_b^{il}}{\sum_{j \in \pi^i}\alpha_b^{ij}}&t_b^i\neq\phi\,.
\end{cases}
\end{align}
Hence, $h_{\alpha z}^k(t_b^i)=\int_0^\infty \Gamma(\kappa_{ki},\theta_{ki})d_{\alpha_b^{ki}}=1$. Similarly for $t_{b-1}^i = \phi$, $\sum_{t_b^i \in \cS_1}h_t(t_b^i)=1$. Therefore \eqref{integral1} always equals $1$ and $h(\bx_b|\bx_{b-1};\bt_b^{ML})$ is a probability distribution function.
\end{proof}
\section{Refinement Full Conditional Distributions}\label{FCDR}
The full conditional distributions for GS in the refinement step can be derived as follows.\\
{ \bf a)} For parent of node $i \in \cN$
\begin{equation}
\begin{aligned}
&f(z_b^i|\bx_{b-1}^m,\bd_{\bB_b},\bz_b^{\overline{i}},\bt_b,\balpha_b) \propto \\
&\begin{cases}
\delta(z_b^i-\phi)& \text{if } {t^i}_{b-1}^m = \phi , t_b^i=\phi \,,\\
\delta(z_b^i-{z^i}_{b-1}^m)& \text{if } {t^i}_{b-1}^m \neq \phi ,
t_b^i\neq\phi \,,\\
f(t_b^i|z_b^i,\alpha_b^{iz_b^i},t_b^{z_b^i}) f(z_b^i|\alpha_b^{ik})_{k
  \in \pi^i}& \text{if } {t^i}_{b-1}^m = \phi , t_b^i\neq\phi \,.\\
\end{cases}
\end{aligned}
\end{equation}
{ \bf b)} For infection time of node $i \in \cN$
\begin{equation}
\begin{aligned}
&f(t_b^i|\bx_{b-1}^m,\bd_{\bB_b},\bz_b,\bt_b^{\overline{i}},\balpha_b) \propto \\
&\begin{cases}
\delta(t_b^i-{t^i}_{b-1}^m)& \text{if } {t^i}_{b-1}^m \neq \phi \,,\\
\begin{tabular}{@{}c@{}}$f(\bd_{\bB_b}^i|t_b^i) f(t_b^i|z_b^i,\alpha_b^{iz_b^i},t_b^{z_b^i})\times$\\
$\prod_{k \in \cC_b^i} f(t_b^k|\alpha_b^{ki},t_b^i)$\end{tabular}&
\text{if } {t^i}_{b-1}^m = \phi \,.\\
\end{cases}
\end{aligned}
\end{equation}
{ \bf c)} For link strength between nodes $i\in \cN$ and $j \in \pi^i$
\begin{equation}
\begin{aligned}
&f(\alpha_b^{ij}|\bx_{b-1}^m,\bd_{\bB_b},\bz_b,\bt_b,\balpha_b^{\overline{ij}}) \propto \\
&\begin{cases}
f(\alpha_b^{ij})& \text{if } {t^i}_{b-1}^m = \phi , t_b^i=\phi \,,\\
\delta(\alpha_b^{ij}-{\alpha^{ij}}_{b-1}^m)& \text{if } {t^i}_{b-1}^m
\neq \phi , t_b^i\neq\phi \,,\\
f(t_b^i|z_b^i,\alpha_b^{iz_b^i},t_b^{z_b^i})f(z_b^i|\balpha_b)
f(\alpha_b^{ij}) & \text{if } {t^i}_{b-1}^m = \phi , t_b^i\neq\phi \,.
\end{cases}
\end{aligned}
\end{equation}
Here ${t^i}_{b-1}^m$, ${z^i}_{b-1}^m$, and ${\alpha^{ij}}_{b-1}^m$
respectively denote the infection time of node $i$, the parent of node
$i$, and the link strength between nodes $i$ and $j$ in the sample
$\bx_{b-1}^m$. 

\bibliographystyle{IEEEtran} 
\bibliography{refs}

\end{document}